\documentclass[twocolumn]{aastex62}
\usepackage{placeins}
\usepackage{siunitx,etoolbox}

\usepackage{hyperref}
\graphicspath{{./}{figures/}}

\received{June 17, 2019}
\revised{\today}
\accepted{}
\submitjournal{AAS}

\shorttitle{K-band Emission in LLAGN}
\shortauthors{Dumont et al.}

\begin{document}

\title{Surprisingly Strong K-band Emission Found in Low Luminosity Active Galactic Nuclei}

\correspondingauthor{Antoine Dumont}
\email{antoine.dumont.neira@hotmail.com}

\author{Antoine Dumont \href{mailto:antoine.dumont.neira@hotmail.com}}
\affil{Department of Physics and Astronomy, University of Utah\\
115 South 1400 East, Salt Lake City, UT 84112, USA}
\author{Anil C. Seth \href{mailto:seth@astro.utah.edu}}
\affiliation{Department of Physics and Astronomy, University of Utah\\
115 South 1400 East, Salt Lake City, UT 84112, USA}

\author{Jay Strader  \href{mailto:strader@pa.msu.edu}}
\affiliation{Department of Physics and Astronomy Michigan State University Biomedical \& Physical Sciences\\567 Wilson Rd, Room 3275 East Lansing, MI 48824-2320}

\author{Jenny E. Greene  \href{mailto:jgreene@astro.princeton.edu}}
\affiliation{Department of Astrophysical Science,Princeton University\\
4 Ivy Lane
Princeton University, Princeton, NJ 08544}

\author{Leonard Burtscher  \href{mailto:burtscher@strw.leidenuniv.nl}}
\affiliation{Leiden Observatory, PO Box 9513, 2300 RA, Leiden, The Netherlands}

\author{Nadine Neumayer  \href{mailto:neumayer@mpia.de}}
\affiliation{Max-Planck-Institut f\"ur Astronomie (MPIA)\\ 
K\"onigstuhl 17, 69121 Heidelberg, Germany}



\begin{abstract}

We examine the near-infrared (NIR) emission from low-luminosity AGNs (LLAGNs).  
Our galaxy sample includes 15 objects with detected 2-10 keV X-ray emission, dynamical black hole mass estimates from the literature, and available Gemini/NIFS integral field spectroscopy (IFU) data. We find evidence for red continuum components at the center of most galaxies, consistent with the hot dust emission seen in higher luminosity AGN.  We decompose the spectral data cubes into a stellar and continuum component, assuming the continuum component comes from thermal emission from hot dust.  We detect nuclear thermal emission in 14 out of 15 objects. This emission causes weaker CO absorption lines and redder continuum ($2.05-2.28\:\mu$m) in our $K$-band data, as expected from hot dust around an AGN. The NIR emission is clearly correlated with the 2–10 keV X-ray flux, with a Spearman coefficient of $r_{spearman}=0.69$ suggesting a $>99\%$ significance of correlation, providing further evidence of an AGN origin.  Our sample has typical X-ray and NIR fluxes $3-4$ orders of magnitude less luminous than previous work studying the NIR emission from AGN. We find that the ratio of NIR to X-ray emission increases towards lower Eddington ratios.  The NIR emission in our sample is often brighter than the X-ray emission, with our $K$ band AGN luminosities comparable to or greater than the 2-10 keV X-ray luminosities in all objects with Eddington ratios below 0.01\%. 
 The nature of this LLAGN NIR emission remains unclear, with one possibility being an increased contribution from jet emission at these low luminosities.  These observations suggest JWST will be a useful tool for detecting the lowest luminosity AGN.  

\end{abstract}


\section{Introduction\label{sec:intro}} 

One of the most important discoveries of the last two decade in astrophysics was the realization that galaxies hold extremely massive black holes in their centers, supermassive black holes (SMBH). These SMBH have masses ranging from million to billion times the mass of the sun and located at the center of the galaxy. 
Observations of inferred SMBH masses ($M_{\bullet}$) show a correlation between them and the mass of the galaxy bulge ($M_{bulge}$) \citep[e.g.][]{Haring2004}, 
as well as SMBH growth rate and star formation of the host galaxy. This suggests that they grew in lockstep. This relation between SMBH and galaxy bulge is expected to come from radiation pressure feedback, is not thought to be caused gravitational influence; the  mass of the SMBH is very small compared to the galaxy bulge ( $M_{\bullet}\sim 0.3\%M_{bulge}$) \citep{McConell2012}.
The radiation is produced by gas and dust that is under the gravitational force of the SMBH, forming a disk of in-falling material (i.e. 'Accretion disk') that gets compressed and heated up due friction, emitting light across the electromagnetic spectrum, and constituting what is called an Active galactic-Nucleus (AGN). AGN activity can give us insights on SMBH growth, SMBH formation history, and SMBH demographics.
\par The Eddington luminosity ($L_{edd}$) sets the maximum radiation pressure that gas in the outer layers of the accretion disk can experience before losing gravitational cohesion and being blown away. \cite{Ho2009} using the Palomar survey show that locally, most massive galaxies have detectable 2-10 keV X-ray emission at their centers, but only a few percent of these accrete at the rates of bright AGN with Eddington ratios $\log(L_{bol}/L_{edd}) >-2$. Thus, a vast majority of galaxies in the local universe are therefore low-luminosity AGNs (LLAGN).
\par
LLAGNs, besides being more numerous, are also observed to have a different Spectral Energy Distribution (SED) than their bright counterparts, showing an excess of emission at radio and infrared wavelengths \citep{Ho2008}, indicating that they are not just a scaled down version of brighter AGN and a different accretion mechanism could be associated with them; when the accretion rate of the SMBH is very small e.g. $L_{bol}\ll L_{edd}$, the optically thick geometrically thin disk is believed to change to a optically thin geometrically thick radiatively inefficient quasi-spherical state called RIAF or radiatively inefficient accretion flow (see \citet{Quataert2001} for a comprehensive review). The accretion disk of an AGN typically peaks at the UV and optical wavelengths. These UV/optical photons according to the AGN unified model \citep{antonucci93,krolik1998} can be scattered-up by inverse-Compton scattering in a hot corona surrounding the ``accretion disk" producing AGN X-ray emission \citep{Haardt1994A}. X-rays have widely been used as a AGN tracer since it suffers from less absorption than UV/optical photons. 
The UV/optical photons that are absorbed by dust surrounding the accretion disk (the so-called torus) are re-emitted at infrared (IR) wavelengths. Synchrotron emission from AGN jets can also contribute across wavelength,  especially for "radio loud" AGN \citep{Yuan2005,Perlman_2007,Mason2012,Prieto2016}. A tight correlation is seen between the AGN emission at mid-IR wavelengths and X-ray wavelengths \citep{Levenson2009,Gandhi2009,Mason2012,Asmus2015}. 
This correlation, together with a similarity of the luminosities, and the presence of a thermal component in the AGN spectra at mid-infrared wavelength with a temperature of $\sim\: 300\:K$ are all well explained by the torus model \citep{Priet2010}. In this scenario, the mid-IR emission is a result of disk reprocessing, and thus shares the accreting central SMBH as origin with the reprocessed X-ray emission.  The mid-IR--X-ray correlation appears to break down in the highest luminosity AGN \citep{Stern2015}, which may be due to a reduction in the size of the corona or a change in the covering fraction of obscuring material \citep{Toba2019}.

\par Less understood is the near-infrared (NIR) emission from AGN. This NIR emission is more affected by obscuration than in the mid-IR, and stellar contamination is a greater challenge.  Early observations showed the presence of bright NIR point sources associated with luminous AGN \citep{Kotilainen1992,Quillen2001,ALONSO-HERRERO-2001}.  This emission may arise at the dust sublimation radius, where dust can be heated up to 2000~K \citep{Laor1993,Kobayashi1993,Alonso-herrero1996,Riffel2009,Kishimoto2011}. 
This NIR emission also seems to be correlated with X-ray emission \citep{Kotilainen1992,Alonso-herrero1996,Civano2012,Burtscher2015,Muller-Sanchez2018}, supporting this dust emission picture.  In lower luminosity systems, the development of NIR integral field spectrographs has enabled spectral separation of stellar and AGN emission  \citep{Davies2007,Riffel2010,Seth2010ngc404,Seth2010m32,Burtscher2015,Muller-Sanchez2018}.  More specifically, these studies have used the dilution of the CO(2,0) bandhead at $2.3\mu$m and the slope of the $K$-band spectra to disentangle the non-stellar NIR emission from AGN activity from stellar (along with some earlier studies with the use of long-slit spectra \citep{Ivanov2000,Imanishi2004}).  The dilution and spectral reddening is found at the galaxy center, as expected from AGN activity.  Furthermore, reverberation and NIR interferometric measurements have shown sizes of $<$0.1~pc for this emission in some nearby bright AGN  \citep{Minezaki2006A,Minezaki2006B,Suganuma2006,Kishimoto2011,Mandal2018,Sturm2018}.  
The geometry of the system can also constrain the physical mechanism of the NIR emission.  The NIR emission has been found to be elongated
along the accretion direction, suggestive of the source being hot dust on the inner edge of the torus \citep{Minezaki2006B,Minezaki2006A,Suganuma2006,Kishimoto2011,Weigelt2012}. On the other hand, resolution of the mid-IR data using interferometric measurements shows elongation along the polar direction, which has been used to suggest the mid-IR emission comes from dust emission in the outflowing wind \citep{Honig2012,LopezGonzaga2016A}, challenging the early assumed torus model.  These studies have all been done on bright, high accretion rate AGN.
 
\par Here, we focus on studying the NIR AGN emission in a sample of nearby LLAGN (we define objects with $L_{bol}/L_{edd} \leq 10^{-3}$ as LLAGN).  Two galaxies in this luminosity range have previously been found to have surprisingly strong NIR emission relative to their X-ray emission \citep{Seth2010ngc404,Seth2010m32}.  However, larger samples of NIR emission have been restricted to AGN with $L_{bol} \gtrsim 10^{41}$~ergs/s \citep{Burtscher2015,Muller-Sanchez2018}.  In this work, we study the NIR AGN emission and its relation to X-ray emission in a large sample of LLAGN.  By using a sample with dynamical BH masses, we also can study the dependence of this NIR emission on accretion rate.  

This paper is organized into 5 sections. In Section \ref{sec:sample_selection} we describe the sample selection and observations. Section \ref{sec:Results} describes our two methods to obtain NIR fluxes, and we present maps of the derived NIR emission. Section \ref{sec:NIRvsXRAY} Compares our NIR fluxes estimates with 2-10 keV X-ray nuclear fluxes, and the relation of NIR fluxes with Eddington ratio. In Section \ref{sec:non-agn} we discuss some other possible explanations for the origin of the thermal NIR fluxes. Finally, Section \ref{sec:conclusion} we discuss our results and conclude.

\begin{deluxetable*}{l| L l L L L L C}[h!]
\tablecaption{summary of objects\label{tab:table1}}
\tablehead{
\colhead{Galaxy name} & \colhead{$BH\:Mass$} & \colhead{$BH\:Mass\:method$} &
\colhead{$NED\:distance$} & \colhead{ $2$-$10\:keV\:X$-$ray\:flux$} & \colhead{$Lx\:2$-$10 keV$}& \colhead{$\log_{10}\:(Edd ratio)$}  &\colhead{$AGN$ } \\
\colhead{} & \colhead{} & \colhead{} & \colhead{$Mpc$} &
\colhead{$(erg/s\:cm^{2})$} & \colhead{$(erg/s)$} & \colhead{}&\colhead{$activity$}}
\startdata
M32 &  2.50e+06^{(a)}&   stellar &   0.76 &          1.04e-14^{(g)} &   35.86 &   -7.42 & R^{(r)}\\
M60-UCD1 &    2.10e+06^{(b)} &      stellar &     16.50 &     3.30e-16^{(f)} &     37.03 &    -6.18 & \\
NGC404 & <1.50e+05^{(c)}&  stellar & 3.79 &   2.67e-14^{(j)} &  37.66 &    >-4.40 & L2^{(m,p)} \\
NGC4762 &  5.00e+07^{(d)} &        stellar &              16.26 &          1.92e-15^{(k)} &        37.78 &   -6.82 & L2^{(p)},R^{(q)} \\
NGC4434 &  9.00e+07^{(d)} &        stellar  &              22.74 &          1.70e-15^{(k)} &        38.02 &   -6.82 & \\
NGC821 & 1.86e+08^{(e)}&        stellar &              26.28 & 4.30e-15^{(g)} &        38.55 &   -6.61 & A^{(p)} \\
NGC4736 &  7.00e+06^{(e)}&        stellar &               5.10 &          1.61e-13^{(g)} &        38.70 &   -5.02  & L2^{(m,p)},R^{(q)}\\
NGC4339 &  6.50e+07^{(d)} &        stellar &              16.67 &          1.86e-14^{(g)} &        38.79 &   -5.91  & A^{(p)}\\
NGC4486B &  4.00e+08^{(e)} &        stellar &              15.36 &          3.43e-14^{(g)} &        38.98 &   -6.51 &  \\
NGC4552 &   5.00e+08^{(e)} &  stellar &     16.47 &   9.23e-14^{(i)} &        39.47  &   -6.11  & T^{(p)} R^{(l,q)} \\
NGC4395 &  1.07e+04^{(f)} &        stellar &               4.22 &          3.21e-12^{(g)} &        39.83 &   -1.08 & Sy1.8^{(m,p)}\\
M104 &  7.00e+08^{(e)} &        stellar  &              11.32 &          6.64e-13^{(g)} &        40.00 &   -5.71  & R^{(l)},Sy2^{(m)} \\
NGC4486 &  6.00e+09^{(e)} &        stellar &              16.74 &          9.85e-13^{(g)} &         40.52 &   -6.16 & RJ^{(n,q)},L^{(m)}\\
NGC4258 &  4.00e+07^{(e)}&          maser &               7.27 &          9.06e-12^{(g)} &        40.75 &   -3.71  &L1.9^{(p)} RJ^{(n)}\\
NGC3393 &  2.00e+07^{(e)} &        stellar &              49.20 &          2.26e-13^{(g)} &        40.81 &   -3.27  & Sy2^{(m)}\\
\enddata
\tablecomments{References for Black hole Masses: $^{(a)}$\citet{verolme2002}, $^{(b)}$\citet{Seth2014m60},$^{(c)}$\citet{Nguyen2018},$^{(d)}$\citet{Krajnovic2018},$^{(e)}$\citet{Saglia2016},$^{(f)}$\citet{She2016}, Reference for absorption corrected 2-10 keV X-ray fluxes: $^{(g)}$\citet{She2016},$^{(h)}$\citet{Ahn2018},$^{(j)}$\citet{Binder2011},$^{(k)}$\citet{Gallo2018},$^{(i)}$\citep{Gultekein2019A}, References for AGN activity:$^{(l)}$\citet{Healey2007},$^{(m)}$\citet{Veron2006},$^{(n)}$\citet{Liu2002},$^{(p)}$\citet{Ho1997},$^{(q)}$\citet{Nagar2005},$^{(r)}$\citet{Yang2015} ``L" represents LINER,``S" represents Seyfert, ``T" represents objects with a LINER + H II type region spectrum, and ``A" indicates absorption-line spectrum as detailed in \citet{Ho1997}. ``R" indicates nuclear radio emission, and ``RJ" labels radio jet emission as described in \citet{Liu2002,Nagar2005}.}
\end{deluxetable*}

\section{Data \& Sample Selection} \label{sec:sample_selection}

\par To study the NIR emission from low-luminosity AGN (LLAGN) we focus on adaptive optics integral field data of nearby galaxies.  The high resolution of these observations can help us separate out the AGN and stellar emission from these high surface-brightness nuclei.  

We constructed our galaxy sample shown in Table~\ref{tab:table1} by finding galaxies meeting the following conditions:
\begin{itemize}
\item An available dynamical SMBH mass in the literature \citep{Seth2010ngc404,Seth2010m32,She2016,Saglia2016,Nguyen2018,Krajnovic2018}.
\item An available 2-10 keV X-ray luminosity estimate; this was used then to restrict the sample to only LLAGN with $L_{bol}/L_{edd} \leq 10^{-3}$.  We assumed a ratio of $L_{bol}/L_{2-10 keV} = 16$ from \citet{Ho2008} in calculating the bolometric luminosity. 
\item Data in the Gemini/NIFS archive.  We also tested using data from VLT/SINFONI, but found instrumental effects in the relative flux calibration that prevented simple interpretation of the results.  
\end{itemize}
This resulted in the sample of 15 objects; the summary of their properties of is shown in Table~\ref{tab:table1}. We also include the higher Eddington ratio source NGC4395 to provide context.  At these low Eddington ratios, objects typically do not host any broad lines \citep{Ho2008,Elitzur2014}, and indeed NGC4395 is the only object to host a number of broad lines (although broad H$\alpha$ is also seen in NGC4258).  However, other indicators of AGN activity apart from the X-ray emission are found in most of these galaxies -- these are summarized in the last column of Table~\ref{tab:table1}. The X-ray emission in these galaxies is likely due to AGN activity. 
Although emission from X-ray binaries is a possible contaminant, the level of this contamination (even at the low luminosities) of our sample is expected to be small \citep[e.g. see discussion in][]{Miller2012}.  This is particularly true because the X-ray fluxes considered here are all from observations with the Chandra telescope, whose high resolution \citep[FWHM$\sim$0$\farcs$5][]{Weisskopf2002} significantly reduces the possibility of X-ray binary contamination in these nearby galaxies (median distance of 16.5~Mpc).
We included NGC404 in our final sample even though it only has an SMBH mass upper limit (and thus lower limit on the Eddington ratio), since \cite{Seth2010ngc404} detected strong NIR hot dust emission in this object.   Two other galaxies in this sample have previous NIR hot dust flux measurements; we discuss these in Section \ref{subsec:robustness}.  \citep{Seth2010m32,Burtscher2015}.

\subsection{Data reduction \label{subsec:reduction}}
\par The data reduction for the \textbf{Gemini/NIFS} data was done with a combination of IRAF and IDL scripts as described in \citet{Seth2010ngc404} and pipelined by Walsh, J. {\em private communication}. This pipeline is based on the IRAF packages for Gemini/NIFS \textbf{NIFSEXAMPLES} scripts; individual Gemini/NIFS scripts have been altered to enable variance propagation, and the final individual and combined cubes are created using custom IDL packages.  Each object was corrected for telluric absorption using an AV star that is masked to remove the $Br\gamma$ absorption line at 2.16 $\mu$m.  

\begin{deluxetable*}{l|lllcl}[ht!]
\tablecaption{Gemini/NIFS K-band observations \label{tab:table2}}
\tablecolumns{5}
\tablenum{2}
\tablewidth{0pt}
\tablehead{
\colhead{Object} &
\colhead{Program ID} &
\colhead{UT date } & \colhead{Exp. time} &\colhead{Adaptive optics} &\colhead{PI} \\
\colhead{} & \colhead{} & 
\colhead{} & \colhead{(s)} & \colhead{} & \colhead{}
}
\startdata
M32 & GN-2005B-SV-121 & 2005 Oct. 23 & 8 x 600(s) & NGS & N.A \\
M60-UCD1 & GN-2014A-Q-4 & 2014 Feb. 20 & 4 x 900(s) & LGS & Seth$^{(a)}$\\
            & GN-2014A-Q-4 & 2014 May. 13 & 2 x 900(s) & LGS & Seth$^{(a)}$\\
NGC404 & GN-2008B-Q-74 & 2008 Sept. 21 & 4 X 760(s) & LGS & Seth$^{(b)}$\\
           & GN-2008B-Q-74 & 2008 Sept. 22 & 4 X 760(s) & LGS & Seth$^{(b)}$\\
NGC4762 & GN-2010A-Q-19 & 2010 Feb. 26 & 6 x 600(s) & LGS & Krajnovic$^{(c)}$\\
NGC4434 & GN-2009A-Q-54 & 2009 May 08 & 8 x 600(s) & LGS & Krajnovic$^{(c)}$ \\
NGC821 & GN-2012B-Q-89 & 2012 Dec. 29 & 16 x 90(s) & LGS & Scott \\
NGC4736 & GN-2011A-Q-68 & 2011 May 24 & 4 x 500(s) & LGS & Constantin$^{(d)}$\\
NGC4339 & GN-2010A-Q-19 & 2010 Apr. 24 & 6 x 600 & LGS & Krajnovic$^{(c)}$\\
NGC4486B & GN-2006B-SV-103 & 2007 Jan. 07 & 18 x 85(s) & NGS & Davidge$^{(e)}$ \\
NGC4552 & GN-2013A-Q-58 & 2013 May 30 & 4 x 600(s) & LGS & McConnell\\
NGC4395 & GN-2010A-Q-38 & 2010 Mar. 28 & 4 x600(s) & LGS & Seth$^{(f)}$	\\
M104 & GN-2011A-Q-83 & 2011 Feb. 10 & 6 x 640(s) & NGS & Steiner$^{(g)}$\\
NGC4486 & GN-2008A-Q-12 & 2008 Apr. 16 & 4 x 600(s) & LGS & Richstone\\
NGC4258 & GN-2007A-Q-25 & 2007 Apr. 30 & 10 x 600(s) & NGS & Storchi-Bergmann$^{(h)}$\\
NGC3393 & GN-2012A-Q-2 & 2012 Apr. 11 & 4 x 600(s) & LGS & Wang\\
\enddata
\tablecomments{Adaptive Optics: NGC and LGS stand for ``natural guide star" and ``laser guide star" respectively}. References: $^{(a)}$\citet{Seth2014m60},$^{(c)}$\citet{Krajnovic2018},$^{(d)}$\citet{Constantin2012},$^{(e)}$\citet{Davidage2008},$^{(f)}$\citet{Seth2008},$^{(g)}$\citet{Menezes2013},$^{(h)}$\citet{Winge2009}
\end{deluxetable*}

\subsection{Flux calibration\label{subsec:fluxcal}}

We performed photometry on the telluric calibrator to create a flux calibrated spectra that could be used later to calibrate the galaxy cubes. We make use of the accurate 2MASS $K_s$ magnitudes of our telluric star to obtain a spectral flux calibration for the NIFS data.  More specifically, we multiply our telluric cube by a 9500~K blackbody and take the flux at the mean wavelength of the 2MASS $K_s$ filter (the ''isolambda'') and use the 2MASS magnitude to get a conversion factor from counts to $erg\:s^{-1}\:cm^{-2}\:\AA^{-1}$. The flux calibrated spectra is then applied to the data cubes by multiplying each spatial pixel by the $9500 K$ blackbody curve and the conversion factor.

\par For simplicity, we used the HIP~1123 ($A1V$) telluric taken with NGC~404 for our flux calibration.  We then checked to make sure the photometric calibration did not change significantly from object-to-object.  We did this by taking the 2MASS point-source catalog magnitude of each nucleus, and compared this to synthetic photometry measured from our images by multiplying by the 2MASS filter response function in an aperture of 1.5'' radius.  We found a standard deviation of 16\% between our fluxes and the 2MASS catalog; in practice our errors in fluxes derived are larger than this in all cases, and so the systematic errors from the flux calibration are not significant for our results.  

\subsection{PSF estimation\label{subsec:psf}}

Our Gemini/NIFS data is all taken with adaptive optics (see Table~\ref{tab:table2}).  In these observations, the PSF is often complex, with a nearly diffraction limited core and a halo corresponding roughly to the seeing.  The PSF has already been determined for some of our observations through relatively complicated modeling of HST data or making models of the AGN emission \citep{Seth2010ngc404,Seth2010m32,Gebhardt2011,Seth2014m60,Denbrok15}.  However, a detailed model of the PSF is unnecessary here, as we just hope to estimate the total flux of the NIR AGN emission in each object.  In many AGN, broad lines are used to measure the PSF \citep[e.g.][]{Davies2007}, but these are absent in almost all of our objects.  The AGN continuum emission itself can also provide an estimate of the PSF, since the hot-dust emission is expected to be on sub-pc scales (see introduction), and thus unresolved in our observations.  Indeed, the hot dust emission in M32 and NGC404 was found to roughly match the PSF determined using other methods \citep{Seth2010ngc404,Seth2010m32}, although jet emission could make the emission non-pointlike \citep[e.g.][]{Gebhardt2011}.

We estimate the PSF in each of our objects from our derived hot dust emission maps (Section~\ref{sec:Results}) with a single Gaussian component.  We use a single Gaussian because (1) our measurements are not high enough signal-to-noise to support a more complicated PSF model, and (2) a simple model is all we need to ensure that we are measuring most of the NIR AGN flux.  Our fits find a median Gaussian width $\sigma = 0\farcs096$, with the maximum being $0\farcs179$, corresponding to an FWHM of 0$\farcs$42 arcsec.  The best-fit hot dust Gaussian $\sigma$ for each object is listed in Table~\ref{tab:table3}.  Based on this data we use an aperture of $0\farcs3$ radius to measure our fluxes (Section~\ref{subsec:templatesmethod},~\ref{subsec:annularmethod}), which should contain a majority of the AGN flux. NGC3393 with a FWHM of 0$\farcs$42 is the only object with a FWHM bigger than our 0$\farcs$3 radius aperture. The total best-fit AGN luminosity for NGC3393 is $\log{L_{K}}=40.85\:[erg/s]$ which is $\sim17\%$ higher than the flux listed in Table~\ref{tab:table3}, but for consistency we have opted to keep all our AGN NIR fluxes with the same 0$\farcs$3 radius aperture.  We also note that the AGN NIR emission appears to be significantly more compact than the stellar emission in our galaxies, arguing against a stellar population source for this emission (see Section \ref{subsec:robustness} for more discussion).

\section{Detecting NIR AGN Continuum Emission} \label{sec:Results}
\subsection{Evidence for AGN activity}
\label{subsec:evidence_for_agn_activity}
In this section we discuss how we derive the AGN continuum fluxes in the NIR. The light coming from the central regions of the galaxy contains stellar emission from the underlying stellar population, and in luminous AGN, there is a thermal component from the IR reprocessing of more high energetic photons in the accretion disk \& torus. This thermal emission has temperatures of $ 600-1500 \: K$ given by the sublimation temperature of the dust (see introduction \ref{sec:intro}). Depending on how powerful the AGN is, the thermal emission can dominate the total luminosity and virtually erase the stellar component. This has been used to disentangle the hot dust emission in luminous AGNs. For example, \cite{Burtscher2015} used the dilution of the most prominent stellar feature in the K-band, the CO bandhead at $2.3\:\mu$m to estimate the fraction of the light in the spectra that was thermal from AGN activity. They also used a separate SED decomposition method to estimate the thermal emission, where the spectra were fit by a stellar template plus a blackbody. The hot dust flux was estimated by integrating the blackbody over the bandwidth. They found good agreement between both methods.

\begin{figure}[h!]
\centering
\plotone{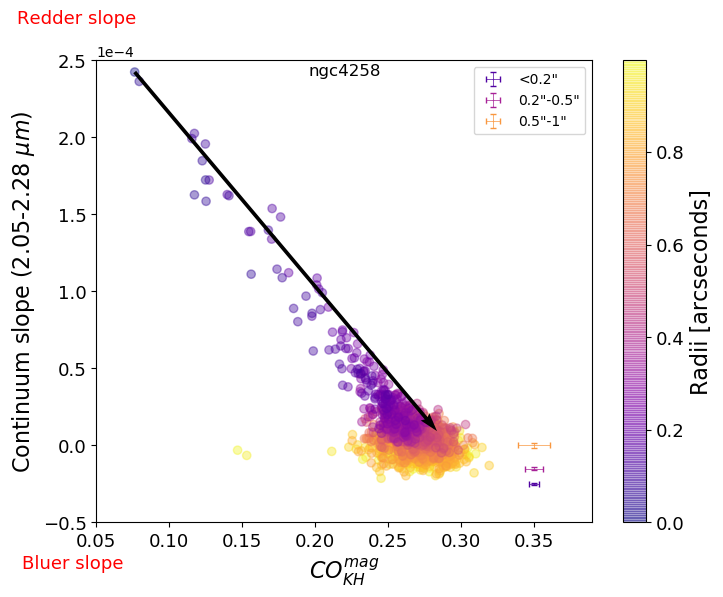}
\plotone{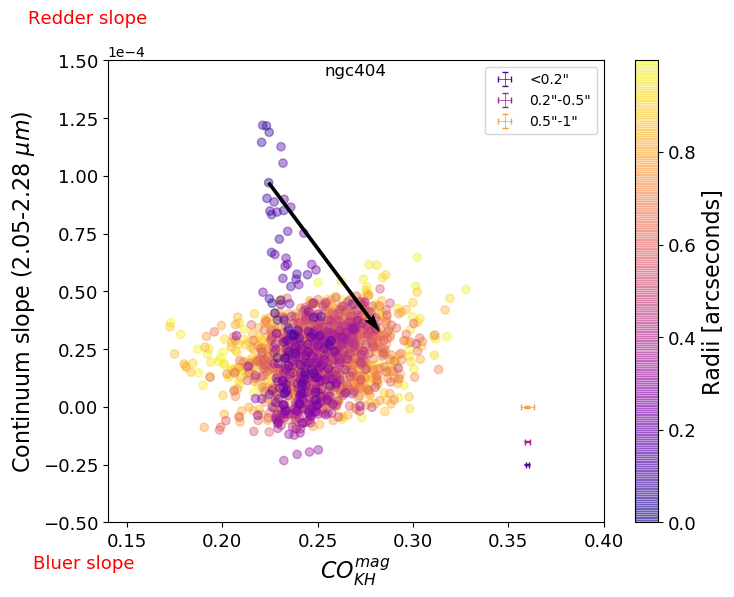}
\caption{The CO bandhead absorption line depth is plotted against the fitted slope of the continuum between ($2.05-2.28\:\mu$m) for each pixel in the IFU map, with larger values indicating redder slopes. Pixels are colored by their radii. The CO bandhead line depth is calculated following the definition in \citet{KH1996}. For stellar photospheres we expect a correlation between these quantities, while the presence of hot dust should redden the slope and weaken the lines. The center of both galaxies (purple pixels) show a deviation towards redder continuua and weaker CO lines as expected for a thermal AGN component.   To quantify this, the arrow shows how the line depth and continuum of  the spectra of the central pixel changes when subtracting the blackbody component from the fit (see figure \ref{fig:fit_example}). The arrow brings back the central pixel to the slope and CO depth values for data at larger radii, consistent with thermal AGN emission see section \ref{subsec:robustness}.}
\label{fig:CO_continuum_plot}
\end{figure}

For LLAGNs the amount of NIR light from AGN activity is small compared to the stellar light of the nuclear region, and we find that the dilution of the CO bandhead is not as prominent. However, we see clear changes in the continuum slope near the centers of many galaxies that are not well fit by stellar templates alone. This effect is clearly observed in Figure~\ref{fig:CO_continuum_plot}, where the CO bandhead absorption line depth is plotted against the fitted continuum slope between ($2.05-2.28\:\mu$m) for each pixel in the IFU data of NGC4258 \& NGC404 as an example. We calculate the CO bandhead absorption line as defined in \citet{KH1996}.
A correlation is expected for stellar photospheres between these two quantities, i.e.~cooler stars have redder slopes and deeper CO bandheads, and hotter stars have bluer slopes and shallower CO bandheads \citep[e.g.][]{Bieging2002}. On the other hand, AGN activity appears as an anti-correlation in the plot displaying redder colors and shallower CO bandhead depths \citep{Seth2010m32}. 

This effect motivates the use of a SED decomposition method to quantify the continuum emission.  We use stellar templates to model the stellar emission, and a set of blackbody template models for the continuum thermal AGN emission. 
For certain objects we found a strong mismatch in the continuum between the stellar templates and the spectrum, thus we used a second method where an annular spectrum from the galaxy itself was used as stellar template.  We describe each method in detail below.

\subsection{Templates Method}\label{subsec:templatesmethod}

We fitted the spectra at each pixel of the Gemini/NIFS data using the Penalized Pixel-Fitting {\tt pPXF} routine \citep{ppxf}. {\tt pPXF} requires a set of stellar templates to model the spectrum of the galaxy. We used a combination of blackbody templates with temperatures running from $600\:K$ to $1500\: K$ to trace the warm \& hot dust plus high resolution synthetic stellar templates from the Phoenix library \citep{phoenixlib} to account for the stellar light.  We used a total of 370 stellar templates from the Phoenix library.  We selected the properties of the stars using Padova isochrones \citep{PARSEC-COLIBRI} with metallicities of $Z_{\odot}=0.019$ and $Z_{\odot}=0.05$ appropriate for typical nuclear star clusters \citep[e.g.][]{Walcher2006,Seth2010ngc404,Kacharov2018} and with stellar ages ranging from $4\:Myr$ to $12.6\:Gyr$ years.  For each individual object a set of templates was selected by fitting the median spectrum within 1" aperture in a wavelength range of  $2.15-2.4\:\mu$m. A night-sky mask from the night-sky spectral atlas \citep{rousselot} was used during the fitting to remove any sky line residuals. \citet{Burtscher2015} used a single stellar template to fit the spectrum of similar $K$-band data and found fluxes that were in good agreement with the fluxes obtained as those derived from dilution of the CO bandhead.

\begin{figure}[h!]
\plotone{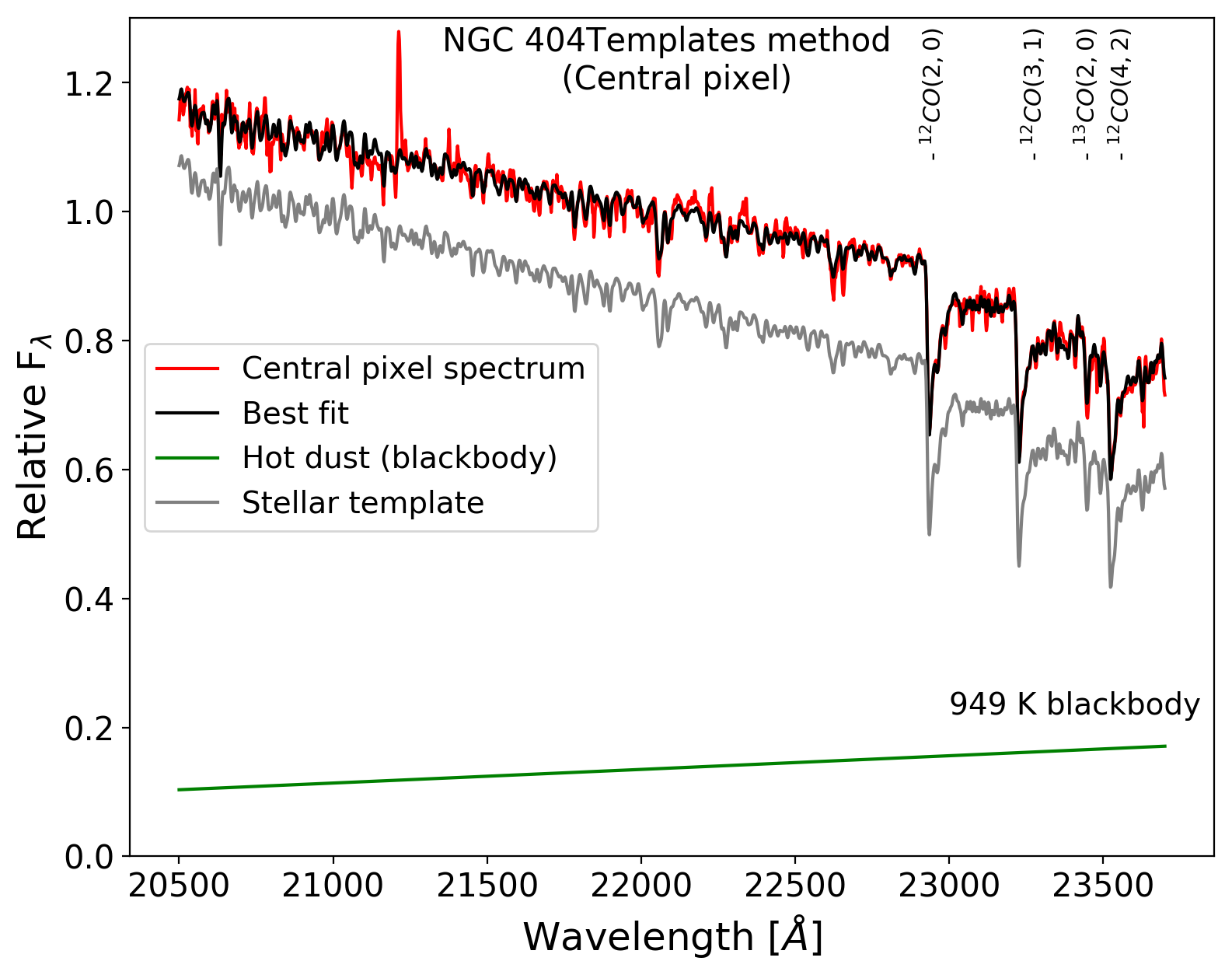}
\caption{Spectral fit of our Gemini/NIFS data in NGC404 to decompose the stellar and AGN continuum emission.  This example shows just the central 0$\farcs$05 pixel spectrum (red), and its decomposition using the stellar template fitting method described in Section~\ref{subsec:templatesmethod}. The y-axis represent the relative flux per unit of wavelength.  The observed spectrum (red) has a shallower continuum slope than the best fit stellar template (grey); this is due to a non stellar thermal emission shown in green. These two components add to create the best-fit model (black).\label{fig:fit_example}}
\end{figure}

\begin{figure*}[h!]
\gridline{\fig{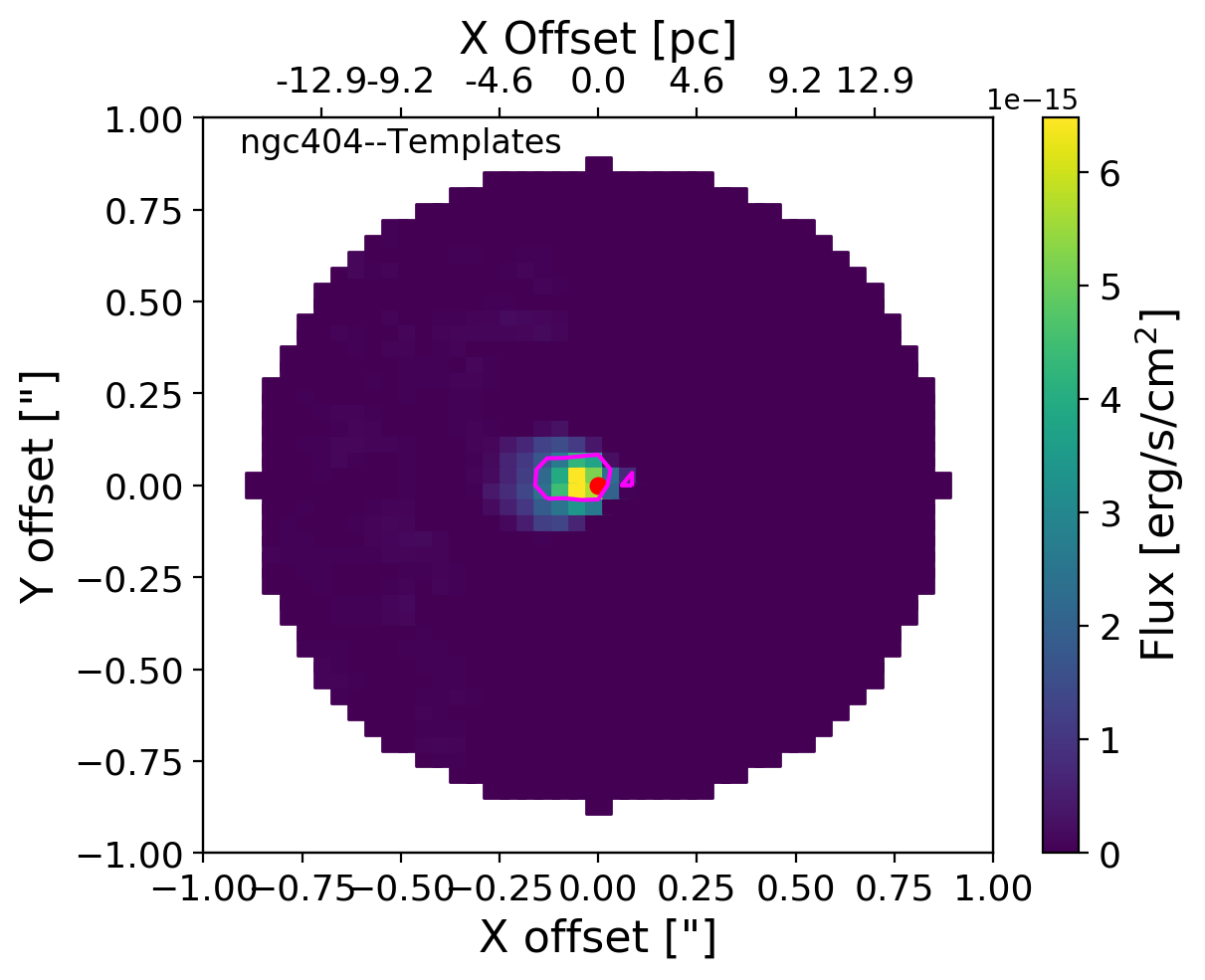}{0.5\textwidth}{}
          \fig{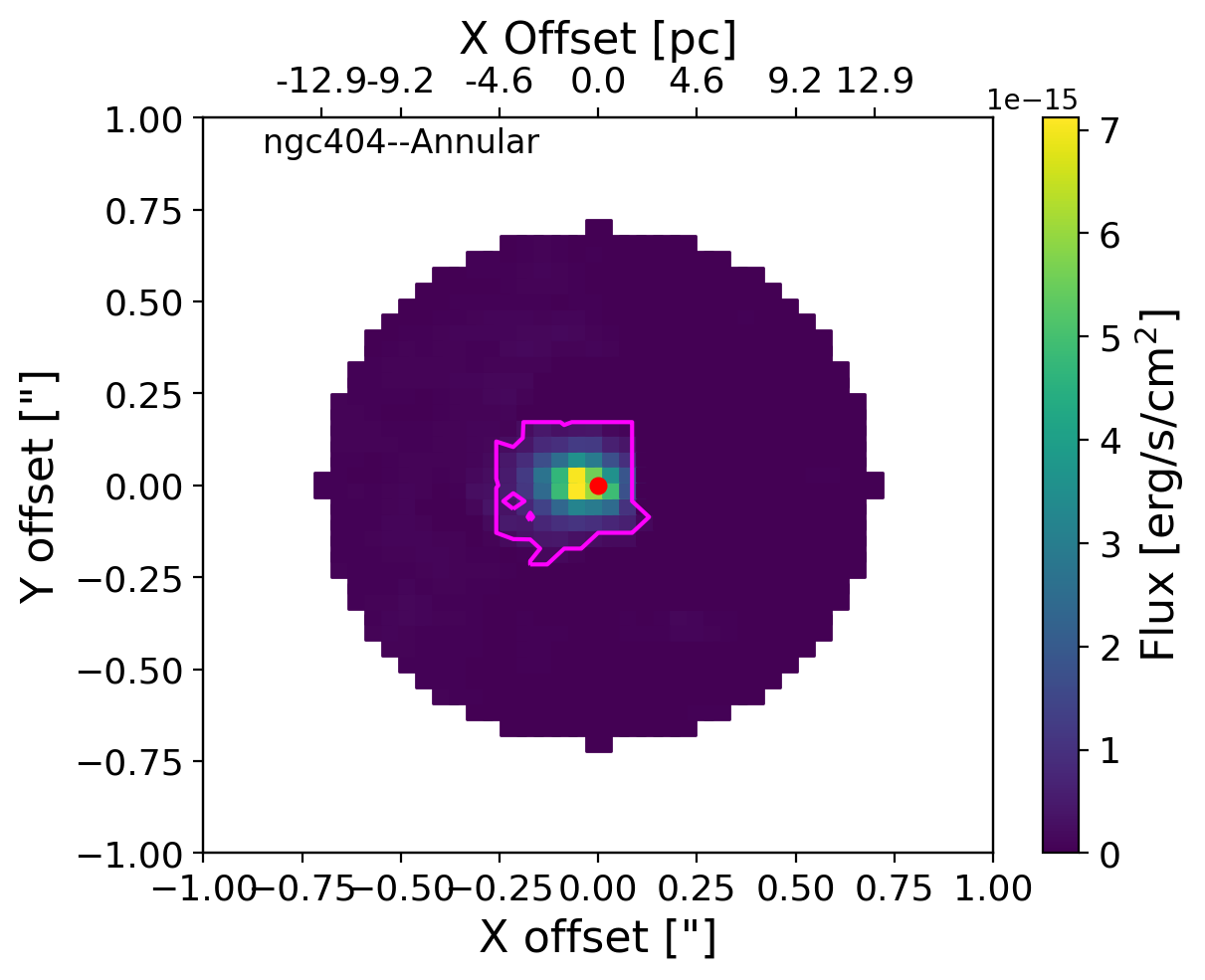}{0.5\textwidth}{}}
          
\gridline{\fig{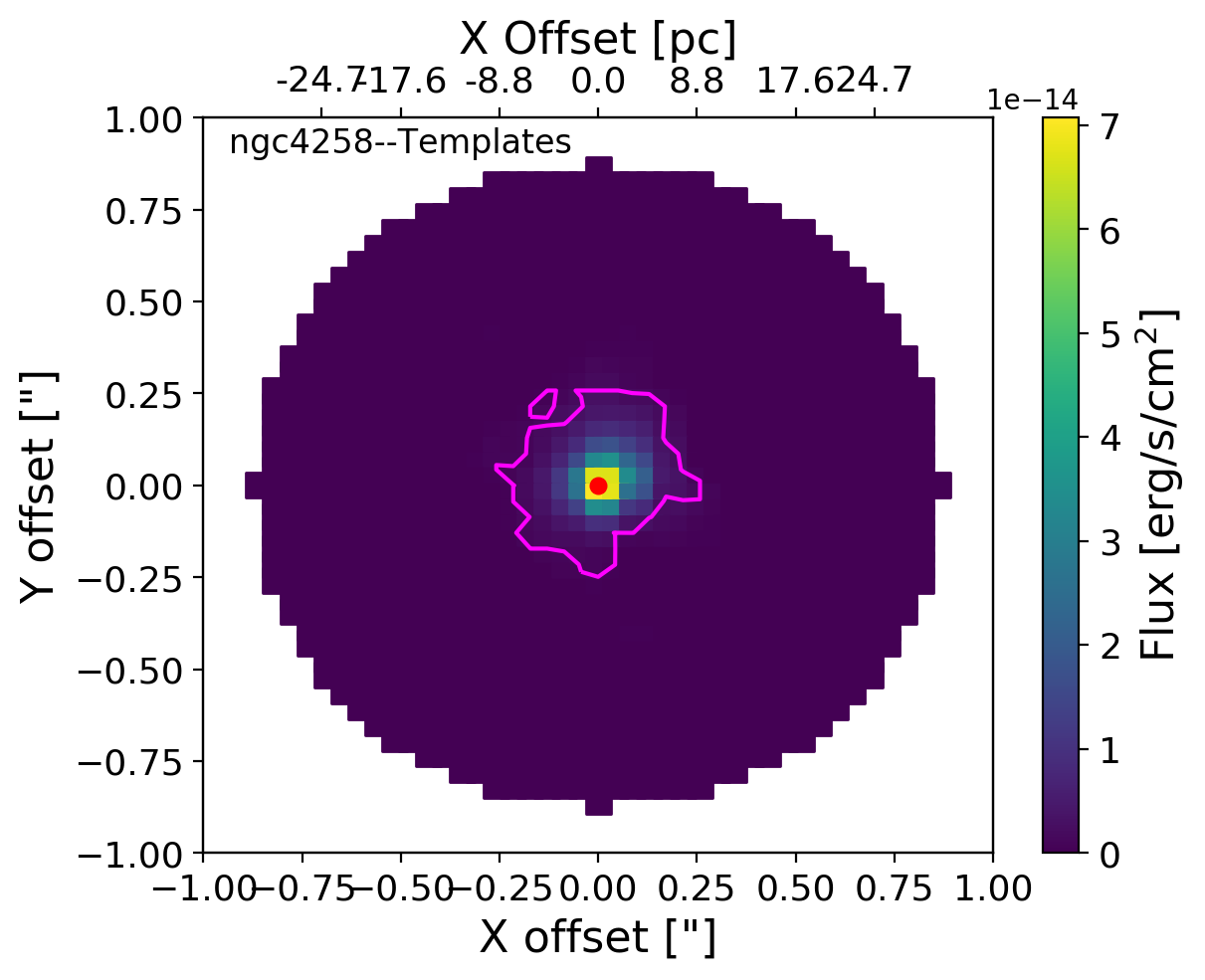}{0.5\textwidth}{(a)}
          \fig{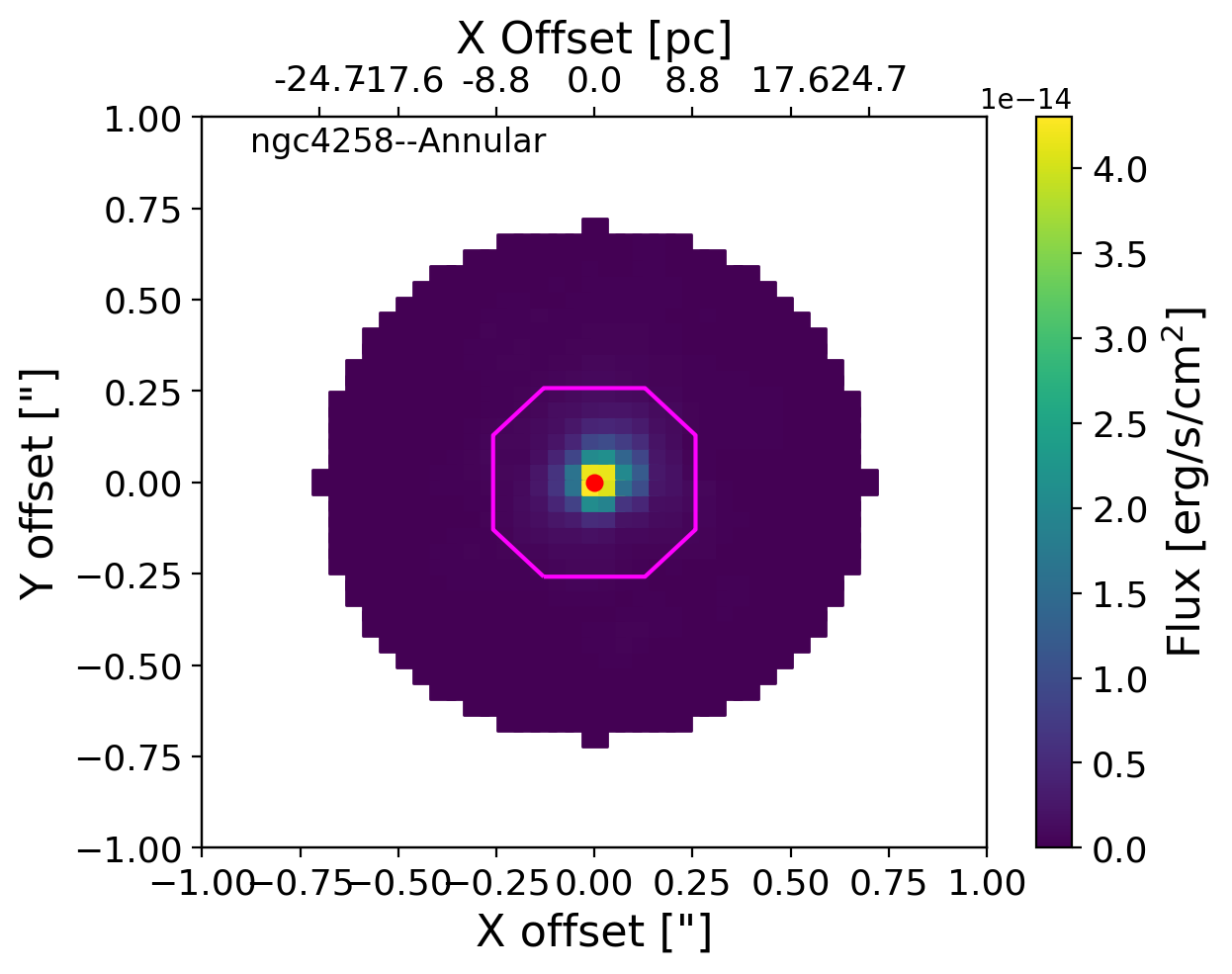}{0.515\textwidth}{(b)}}
\caption{Summary of our hot dust fitting results for two example galaxies.{\em Left \& Right:} The pixels show the flux in the hot dust maps determined using the template method (left), and annular method (right).  The overlaid fuchsia contours represent the pixels where the fluxes are 3$\times$ above the error (see Section~\ref{subsec:templatesmethod} and \ref{subsec:annularmethod}). The color indicates the hot dust flux in erg\,s$^{-1}$\,cm$^{-2}$ at each pixel; this was obtained by integrating the best-fit blackbody between $1.99-2.43\:\mu$m. \label{fig:maps}}
\end{figure*}

An example of the fit for the central pixel of NGC404 is shown in Figure \ref{fig:fit_example}. The observed spectrum (red) is modeled with the combination of a mix of stellar templates (grey) and a mix of blackbody templates with an equivalent temperature of $949\:K$(green). Typically in {\tt pPXF} a polynomial term is used during the fitting process to account for differences in the continuum between the templates and the data.  However, here we have set this polynomial term to zero to enable us to use the continuum slope to model the thermal AGN emission using our blackbody templates.  We fit each spatial pixel in the data cube to create a map, such as the one shown in the left-hand panel in Fig.~\ref{fig:maps}.

We find very small formal uncertainties in these hot dust measurements.  We performed Monte Carlo simulations based on the variance images obtained during the reduction process described in \ref{subsec:reduction}.  The uncertainties in individual pixel fluxes were very small, of order 2\%.  Given this very small formal uncertainty, we expect that template mismatch and our assumed models dominate the uncertainty in our measurements.  To account for these systematic errors in our fitting, we fitted the spectra at each pixel using a subset of the templates; (1) We fixed the blackbody temperature to the minimum or maximum temperature possible ($600\:K$ or $1500\:K$) using a single blackbody template, repeated the fitting and saved flux on a list. (2) We fixed the stellar population by using a single stellar template, repeated the fitting using all the blackbody templates and saved the flux on a list. Thus, at each pixel we get a list of possible hot dust fluxes, with the standard deviation providing an estimate of the systematical error in our fluxes. The systematic errors are of the order of 20\% and thus are much greater than statistical errors.  We quote these systematic errors in our tables and on our plots.

The left-hand panel of Figure \ref{fig:maps} shows a map of the hot dust fluxes in each pixel of the data cube in two of our galaxies (similar plots are available for all galaxies in the Appendix). The flux at each pixel was obtained by the best-fit blackbodies over the bandwidth $1.99-2.43\:\mu$m; fuchsia contours enclose the pixels where $S/N > 3$, i.e.~where the systematic error determined above was smaller than 33\% of the mean best fit flux. The K-band ($1.99-2.43\:\mu$m) hot dust luminosities are shown in table \ref{tab:table3}. These fluxes are calculated over the inner 0$\farcs$3 radius.

\subsection{Annular Method}\label{subsec:annularmethod}

For some objects, we found the templates did not provide a good fit to the stellar spectra even at large distances from the center.  We therefore also did an SED decomposition of the spectra using a stellar template obtained from an annular radius of 0".5 for each galaxy. The stellar template in this case will include any uniform errors in the flux calibration or sky subtraction across the NIFS field of view.  This allowed us to get a better fitting for a few objects that the synthetic stellar templates did not give a good fit. This method assumes that the stellar population in the center is the same as in the annular template and the difference in the slope of the continuum arise from thermal emission.  This method was previously used by \citep{Seth2010m32,Seth2010ngc404} to determine the thermal emission in M32 and NGC404.

As in the templates method, we run Monte Carlo simulations to account for statistical errors and systematic errors.  However, in this case, for our systematic error determination  we can only vary the blackbody templates, not the stellar templates.  As in the templates method, the formal statistical errors in the hot dust fluxes are very small, and the we use the standard deviation of the fluxes fit using varying blackbody temperatures as our error.  The map determined using this method in two objects is shown in right-hand panel of Figure~\ref{fig:maps}. The final fluxes are  given in Table~\ref{tab:table3} for a 0$\farcs$3 radius aperture around the center.

\subsection{Robustness of Detection and Final NIR Fluxes}\label{subsec:robustness}

\begin{figure}
    \centering
    \epsscale{1.15}
    \plotone{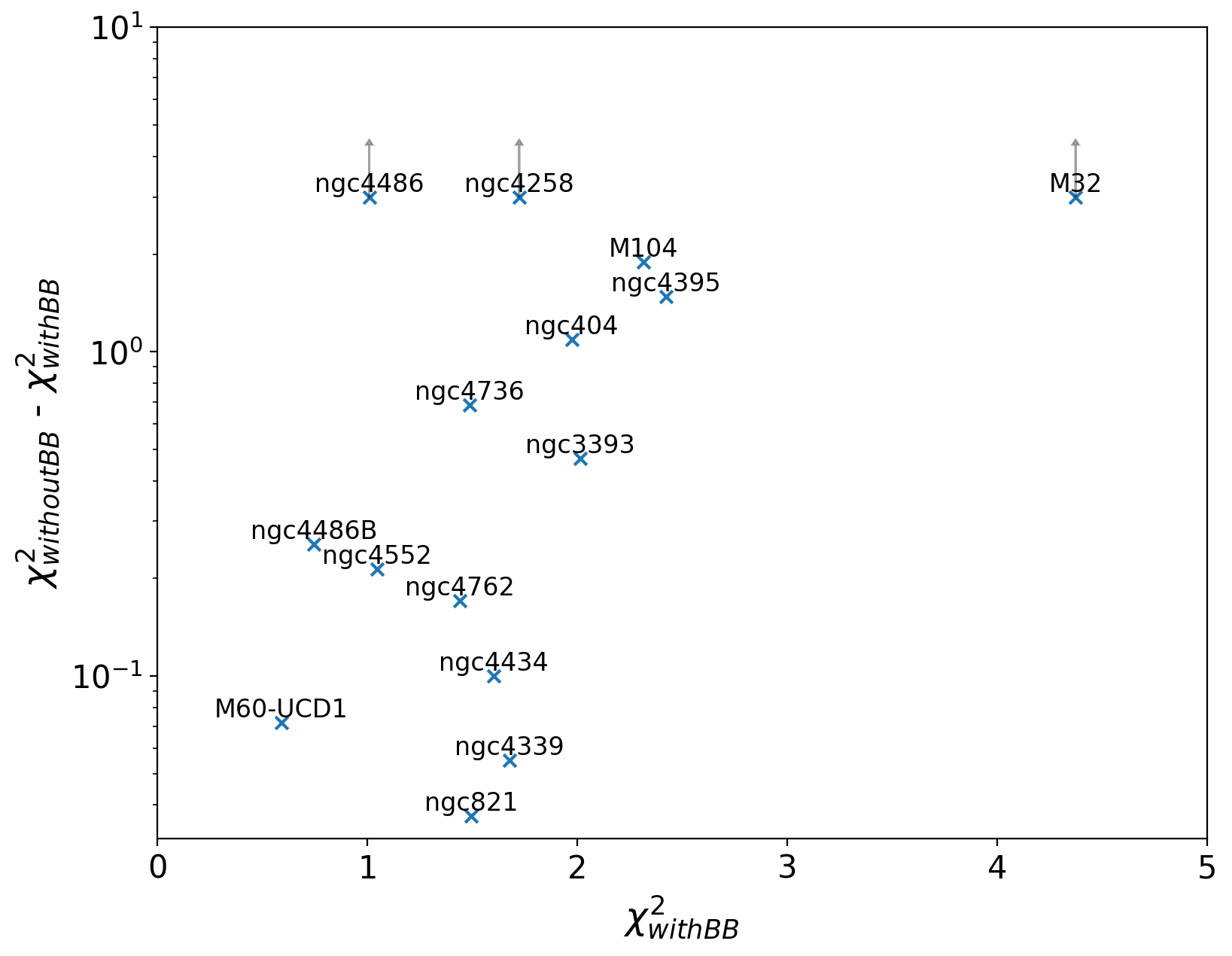}
    \caption{Comparison of goodness-of-fit for the best fitting method with and without blackbody templates. The y-axis shows the difference between the average reduced $\chi^{2}$ over the inner 0$\farcs$3 with and without blackbody templates versus the average reduced $\chi^{2}$ over the inner 0$\farcs$3 with blackbody templates.  In all the objects we obtain a better fit when including blackbody templates.For our number of degrees of freedom ($\sim$1500 per spectra, 60 independent spectra in our 0$\farcs$3 aperture), the difference in average reduced $\chi^2$ for an individual spectrum corresponding to a 1/3$\sigma$ difference is 0.005/0.013, thus the model with a blackbody component is preferred in all the fits. For M32, NGC4486 \& NGC4258 the average reduced $\chi^{2}$ without blackbody templates are $8.75$, $13.7$ \& $59.3$ respectively and they are shown with a vertical arrow for convenience.
    \label{fig:chi2nobb}}
\end{figure}

Our fits thus far have been model-dependent, assuming the presence of a hot dust component.  This is motivated by the fact that in many galaxies, stellar or annular templates alone do a poor job of fitting the central spectra; the slope of the central spectrum is too red . Figure~\ref{fig:chi2nobb} shows  the goodness-of-fit for the inner 0$\farcs$3 with and without including the blackbody templates. In all cases a better fit is obtained when including a blackbody template.

\begin{figure}
\epsscale{1.15}
    \centering
    \plotone{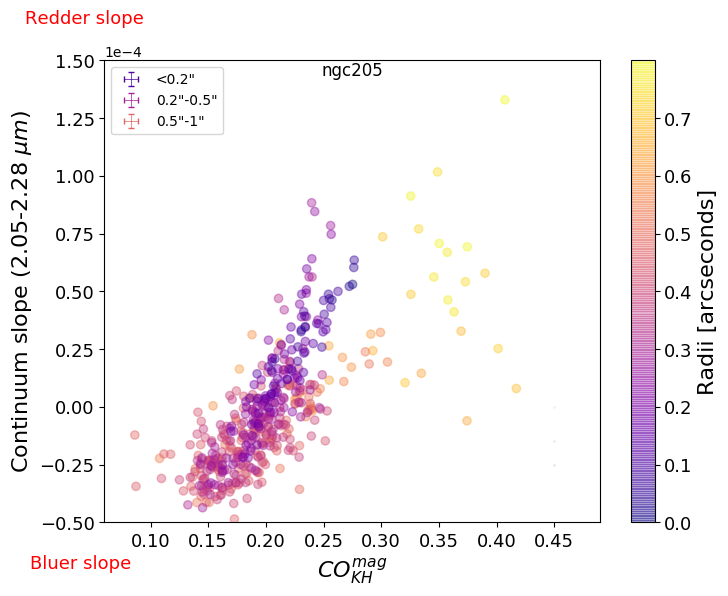}
    \caption{The CO bandhead absorption line depth of NGC205 is plotted against the fitted slope of the continuum between ($2.05-2.28\:\mu$m) for each pixel in the IFU map, with larger values indicating redder slopes.  Pixels are colored by their radii.  No Anti-correlation associated with AGN activity is observed, for more details see \ref{fig:CO_continuum_plot}. }
    \label{fig:ngc205}
\end{figure}

As described in section \ref{subsec:evidence_for_agn_activity} if this emission is due to AGN continuum from hot dust emission (or any other continuum source), the redder slopes are also associated with a dilution in the depth of the CO bandhead absorption lines.
We can therefore assess the robustness of the presence of thermal emission based on whether or not we observe an anti-correlation the continuum vs.~CO bandhead depth plots shown in Figure~{\ref{fig:CO_continuum_plot}}. To help guide our understanding, we also take the best-fit model of the central pixel and observe how the slope and CO bandhead depth changes after subtracting off the best-fit blackbody component.  This is shown as an arrow in Figure \ref{fig:CO_continuum_plot}.  We use the presence of this anti-correlation to test the robustness of our detections; this information is included in  Table~\ref{tab:table3}.  Of the 15 galaxies, 14 were found to be robust detections, with only the detection in NGC~821 not showing a clear anti-correlation signature in the continuum vs.~CO bandhead depth plots.
We also show the case of NGC205 in figure~\ref{fig:ngc205}, where we find no  evidence of AGN activity. NGC205 is not in our sample because despite dynamical evidence for a $\sim$10$^4$~M$_\odot$ BH, the upper limit on the X-ray emission is $<$3$\times$10$^{35}$ ergs~s$^{-1}$ \citep{Dieu2019}. Our non-detection is therefore consistent with these observations.

\begin{figure}[h!]
\epsscale{1.15}
\plotone{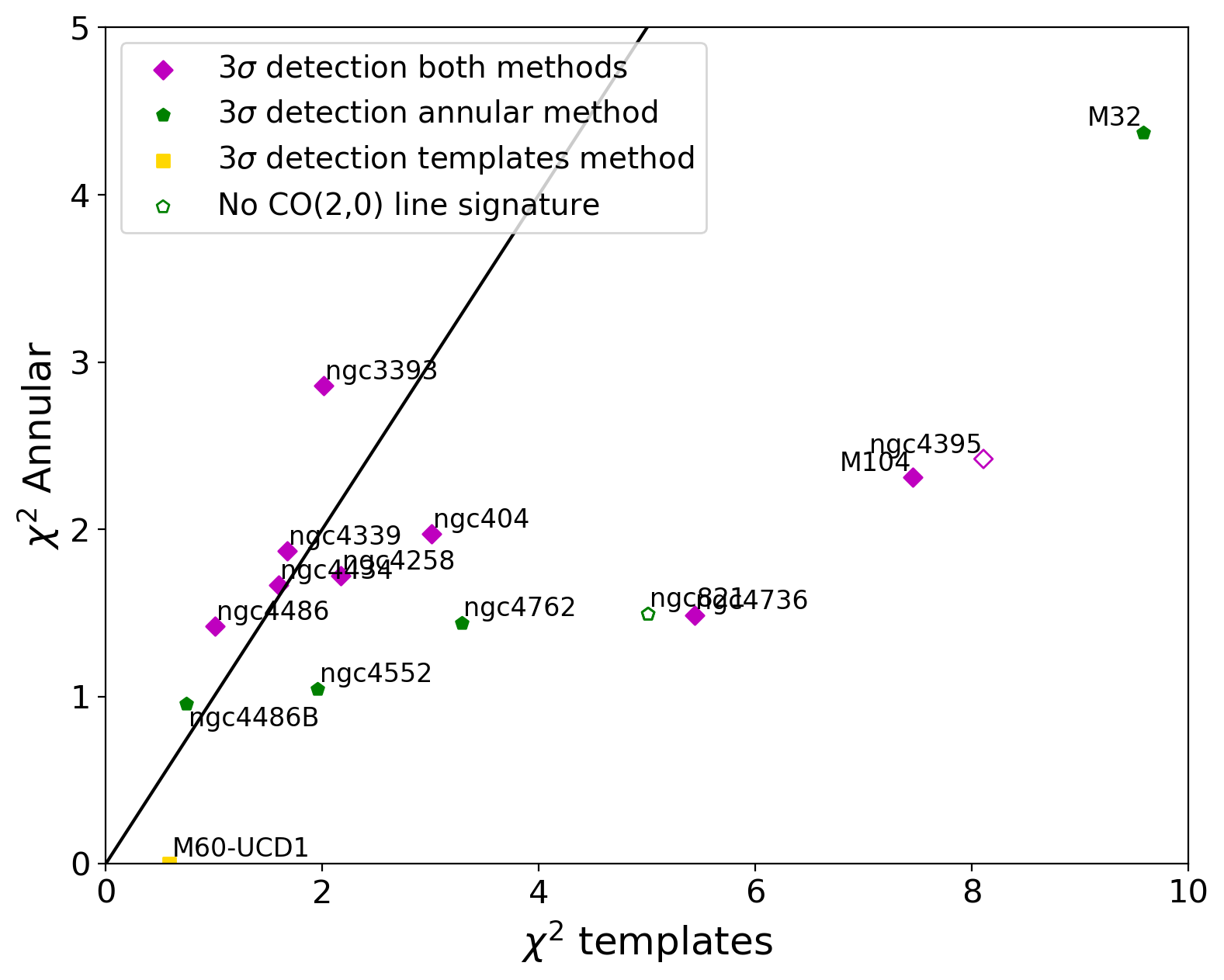}
\caption{Comparison of the goodness-of-fit for the two hot dust fitting methods used in this paper. The axes shows the average reduced $\chi^{2}$ for the inner 0$\farcs$3 for both the annular and template methods described in Section~\ref{sec:Results}.  For our final results, we use the best-fitting method for each object.  The significance of a detection is quantified by taking the standard deviation of the fluxes from the systematic error tests described in Section \ref{subsec:templatesmethod}.  We consider objects with fluxes $>$3$\times$ this standard deviation as robust detection, but in some cases these 3$\sigma$ detections were made only in one of the two methods as indicated by the plot symbols.  The hollow marker indicates an object that doesn't show a clear hot dust CO bandhead-color anti-correlation (as in the right panels of Fig.~\ref{fig:maps}). Due to the compact size of M60-UCD1, no annular method was possible, therefore we have placed it along the templates axis. The values of the average reduced $\chi^{2}$ for NGC4395 are 2.4 \& 8.7 for Annular \& Templates method respectively and are not shown in the figure. This AGN/hot dust dominated galaxy shows no visible CO bandhead absorption \citep{Denbrok15}, 
and thus neither of our fitting methods achieve good fits. \label{fig:annularvstemplateschi2}}
\end{figure}

\begin{figure}[h!]
\epsscale{1.15}
\plotone{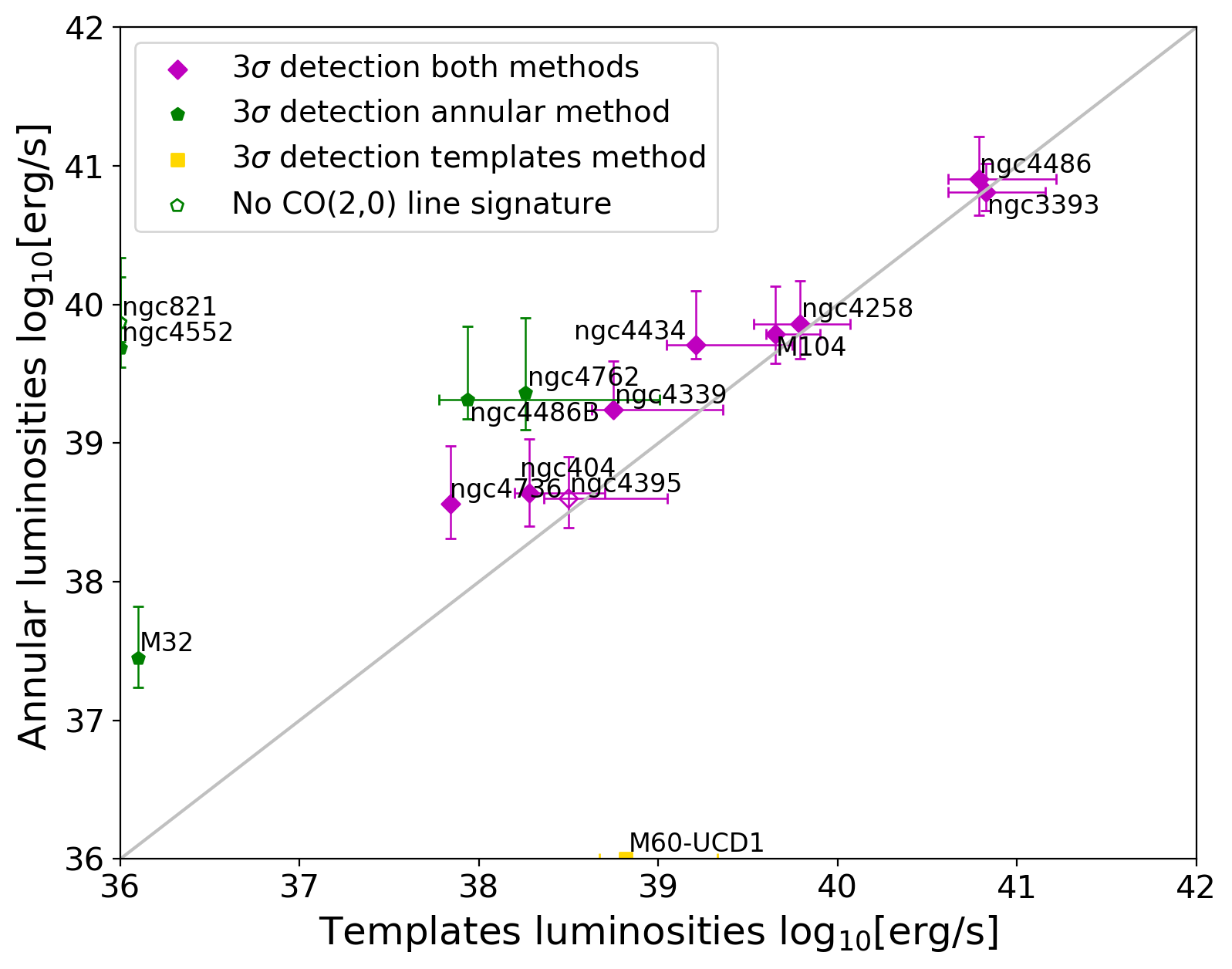}
\caption{Comparison of hot dust luminosities obtained with the template and annular fitting methods. The luminosities were obtained by integrating the best-fit blackbody between $1.99-2.43\:\mu$m. Our errors are dominated by systematics in the fitting (Section~\ref{subsec:templatesmethod}).  The errors shown here are based on the standard deviation of the fluxes obtained during permutations of the templates and blackbodies. The data points are labeled as described in Fig.~\ref{fig:annularvstemplateschi2}.   The objects on the axes (e.g. M60-UCD1,NGC821 \& NGC4552) only have fluxes  determined using one of the two methods.\label{fig:anularvstemplate}}
\end{figure}

To decide which method provided a better estimate of the thermal AGN flux we show a goodness-of-fit comparison in Figure~\ref{fig:annularvstemplateschi2}.  The reduced $\chi^{2}$ values shown are averaged over the inner 0$\farcs$3.  In general, the annular method provides better fits, however this method was not possible in M60-UCD1 due to the lack of flux at large radii, and the template method provided better fits in a handful of objects. In the case of NGC821 \& NGC4552 the templates method was not able to produce a good fit to the spectra, due to strong mismatch in the continuum, and a hot dust NIR flux estimate was possible only with the annular method, where this difference was not present (Figure~\ref{fig:ngc821_comparison}).  Figure~\ref{fig:anularvstemplate} shows that the fluxes between the objects are typically consistent between the two methods, especially in the cases where both methods provide robust estimates.  For each galaxy we use the final fluxes from the method that provides a better reduced $\chi^2$, and list this final flux in bold in Table~\ref{tab:table3}. Also listed in Table~\ref{tab:table3} is the fraction of light in hot dust within our 0$\farcs$3 radius aperture (i.e. the hot dust-to-total flux ratio). The median hot dust fraction is just 0.03 (with a full range of 0.02 to 0.62); this shows that despite the high luminosities we find in the dust in these galaxies, they represent just a small fraction of the nuclear emission in almost all cases. As another check on our fluxes obtained from our pixel-by-pixel analysis, we derived the thermal AGN flux for the summed spectrum in an aperture of 0$\farcs$3. We find a small bias towards smaller fluxes with a median difference of $0.3\:dex$.  This bias is within the typical flux errors for our best-fit hot dust NIR AGN fluxes, and has a minimal affect on the relation between the NIR and nuclear 2-10 keV X-ray luminosity and the ratio of these fluxes to the Eddington ratio that we present in Section~\ref{sec:NIRvsXRAY}.  The small hot dust fractions even in the 0$\farcs$3 aperture explain why hot dust emission has not been found from seeing limited spectroscopic data in similar LLAGN.  For instance, the hot dust fraction in NGC~4258 is 0.22 (22\%) for our 0$\farcs$3 radius aperture, but for a 1$\arcsec$ radius aperture, the fraction of light coming from the hot dust emission drops to just 6\% making it significantly more challenging to spectrally decompose. 

Three objects in our sample have previous NIR ($K$-band) flux estimates. NIR luminosities from \citet{Seth2010ngc404} and \citet{Seth2010m32} for NGC404 and M32 are in good agreement with our luminosities within our errors; this is unsurprising given the identical data and similar methods used. A more independent estimate is available for NGC4486 from \citet{Burtscher2015}; assuming a distance of 16.7~Mpc, their NIR luminsoity is $\log_{10}(L_{NIR})=41.28$; this is slightly above the 1$\sigma$ errors we estimate, however we note that this estimate was made using a larger aperture  and this galaxy is  known to host extended jet emission \citep{Gebhardt2011}.  Overall, our measurements appear to be consistent with previous estimates.

\begin{deluxetable*}{l | C C C L C C C C C}[h!]
\tablecaption{NIR luminosities \label{tab:table3}}
\tablewidth{0pt}
\tablehead{
\colhead{Galaxy name} & \colhead{$\log(L_{K})\:(Templates)$} & \colhead{$\log(L_{K})\:(Annular)$} & \colhead{$Clear\:AGN$}& \colhead{$T_{dust}$} &\colhead{$\log(L_{bol})$} &\colhead{$Required\: A_{v}$} &\colhead{$\sigma_{PSF}$} &\colhead{$Hot\:dust$}&\colhead{$Best$-$fit$} \\
\colhead{} & \colhead{$(erg/s)$} & \colhead{$(erg/s)$} &\colhead{$signature$} &\colhead{$(K)$}&\colhead{$(erg/s)$} &\colhead{$mag$} & \colhead{$\arcsec$}&\colhead{$fraction$}&\colhead{$method$}\\
\colhead{$(1)$} & \colhead{$(2)$} & \colhead{$(3)$} &\colhead{$(4)$}&\colhead{$(5)$}& \colhead{$(6)$}&\colhead{$(7)$}&\colhead{$(8)$} &\colhead{$(9)$}&\colhead{$(10)$}}
\startdata
M32 &  36.10^{0.37}_{0.21} &  \mathbf{37.45^{0.37}_{0.21}} & Yes & 800^{23}_{23} & 38.78^{0.05}_{0.05} & 2.2 & 0.127 & 0.03 &A\\ 
M60-UCD1 &  \mathbf{ 38.82_{0.15}^{0.51} }&  - & Yes & 1099^{127}_{127} & 39.79^{0.19}_{0.22} & 5.3 &0.081 & 0.07 &T\\ 
ngc404 &   38.28^{0.42}_{0.08} & \mathbf{ 38.64^{0.39}_{0.24}} & Yes& 949^{160}_{160} & 39.74^{0.28}_{0.34} &4.4 &0.068 & 0.07 &A\\
ngc4762 &  38.26^{1.25}_{0.09} & \mathbf{ 39.36^{0.54}_{0.27}} & Yes & 615^{67}_{67} & 41.24^{0.20}_{0.23} & 2.9 &0.076 & 0.02 &A\\  
ngc4434 & \mathbf{39.20^{0.54}_{0.16} }&  39.71^{0.39}_{0.10} & Yes & 900^{50}_{50} & 40.31^{0.09}_{0.10} & 3.4 &0.084 & 0.03 &T\\
ngc821 &   - &  \mathbf{ 39.86^{0.47}_{0.14} } & No & 944^{148}_{148} &  40.98^{0.26}_{0.31} & 1.7 &0.111 & 0.02 &A\\ 
ngc4736 &   37.84^{0.29}_{0.10} & \mathbf{ 38.56^{0.42}_{0.25}} & Yes & 606^{32}_{32} & 40.47^{0.10}_{0.11} &4.3 &0.078 & 0.02&A\\  
ngc4339 & \mathbf{ 38.74^{0.61}_{0.21} }&  39.24^{0.35}_{0.01} & Yes & 1033^{33}_{33} & 39.77^{0.05}_{0.05} & 2.6 &0.078 & 0.02 &T\\
ngc4486B &   37.93^{1.07}_{0.16} &  \mathbf{ 39.31^{0.53}_{0.14}} & Yes & 869^{180}_{180} & 40.53^{0.34}_{0.44} &0 .9 &0.134 & 0.03&A\\
ngc4552 &  - & \mathbf{ 39.68^{0.51}_{0.14}} & Yes & 929^{140}_{140} & 40.82^{0.25}_{0.30} &1.9 & 0.129 & 0.03&A\\
ngc4395 & 38.50^{0.55}_{0.14} & \mathbf{ 38.60^{0.30}_{0.21} }& Yes & 718^{118}_{118} & 40.13^{0.29}_{0.35} & 7.0 & 0.057& 0.29 &A\\
M104 & 39.65^{0.25}_{0.05} & \mathbf{ 39.78^{0.35}_{0.21}} & Yes & 801^{83}_{83} & 41.12^{0.10}_{0.21} & 2.0 & 0.107& 0.03 &A\\ 
ngc4486 & \mathbf{40.78^{0.43}_{0.17}} &  40.90^{0.26}_{0.31} & Yes & 1458^{60}_{60} & 41.74^{0.07}_{0.07} &13.7 &0.096 & 0.62 &T\\
ngc4258 & 39.79^{0.28}_{0.26} & \mathbf{ 39.86^{0.25}_{0.31} }& Yes & 1351^{139}_{139} & 40.72^{0.17}_{0.19} & 20.7 &0.061 & 0.22 &A\\
ngc3393 & \mathbf{ 40.71^{0.33}_{0.21}} & 40.81^{0.21}_{0.13} & Yes & 1477^{39}_{39} & 41.66^{0.04}_{0.04} & 5.8 & 0.179& 0.20 &T\\
\enddata
\tablecomments{Summary of NIR hot dust AGN luminosities. Bold luminosities correspond to our best-fit estimation. Column (8) shows the best-fit method with a ``T" or ``A" corresponding to Templates or Annular method resp. .  Column (2) and (3) refer to the total luminosity from 1.99-2.43$\mu$m.  Column (4) ``Yes" indicates the galaxy has a clear CO-bandhead -- color anti-correlation as expected for an AGN.  Column (6) is the bolometric luminosity of the hot dust component for the best-fit temperature and $K$-band luminosity. Column (7) corresponds to the necessary $A_{v}$ extinction that is required to produce the reddening of the data to mimic the change in the slope of the continuum within the central 0$\farcs$1 of each galaxy (see Section~\ref{sec:non-agn}). Column (8) shows the width ($\sigma_{PSF}$) of the best-fit Gaussian function used to estimate the PSF, as explained in Section~\ref{subsec:psf} .Column (9) shows the fraction of the total luminosity in a 0$\farcs$3 radius aperture coming from hot dust emission.}
\end{deluxetable*}

\section{LLAGN in the NIR and their Relation with Bolometric Luminosity} \label{sec:NIRvsXRAY}

\begin{figure*}[h!]
\plottwo{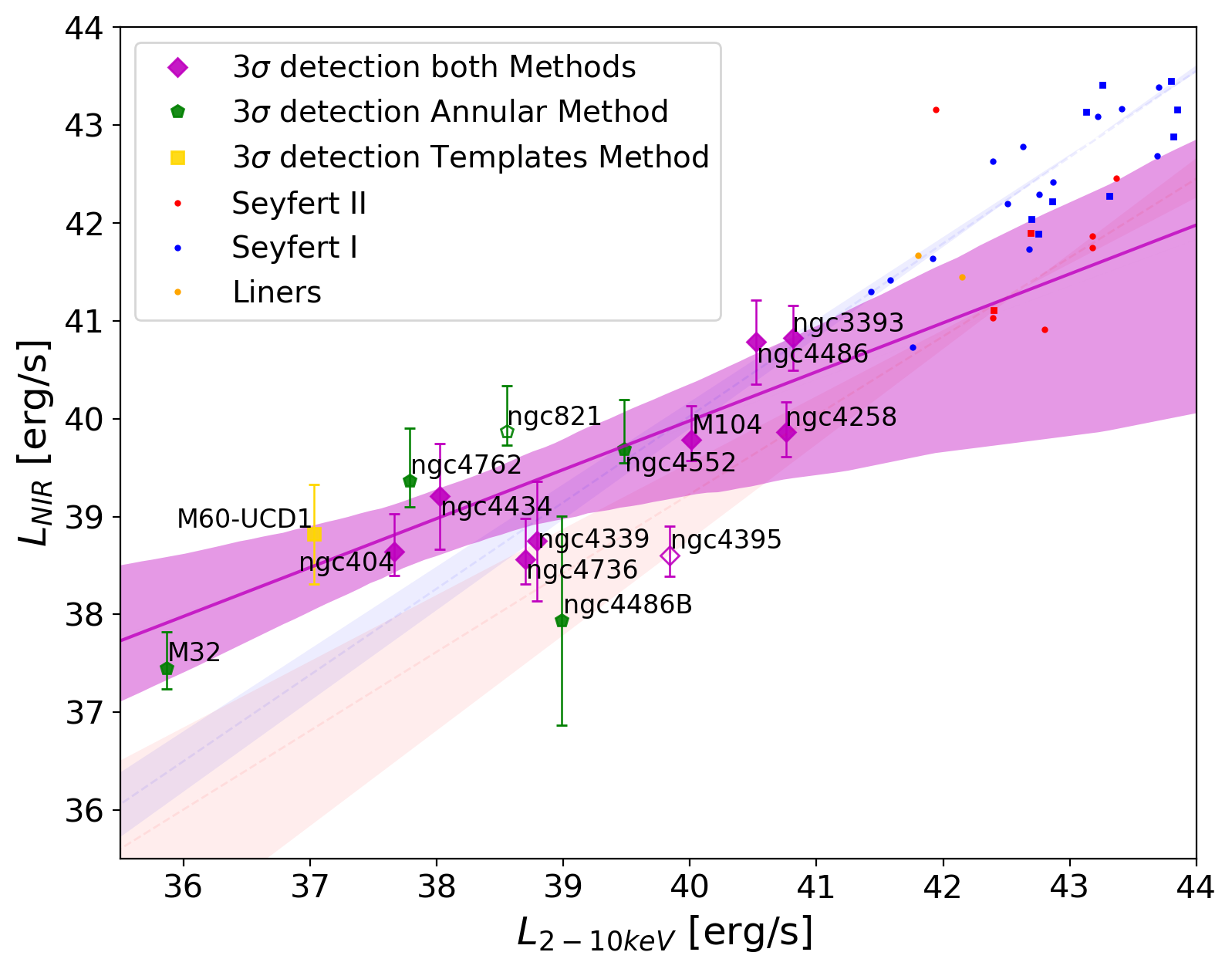}{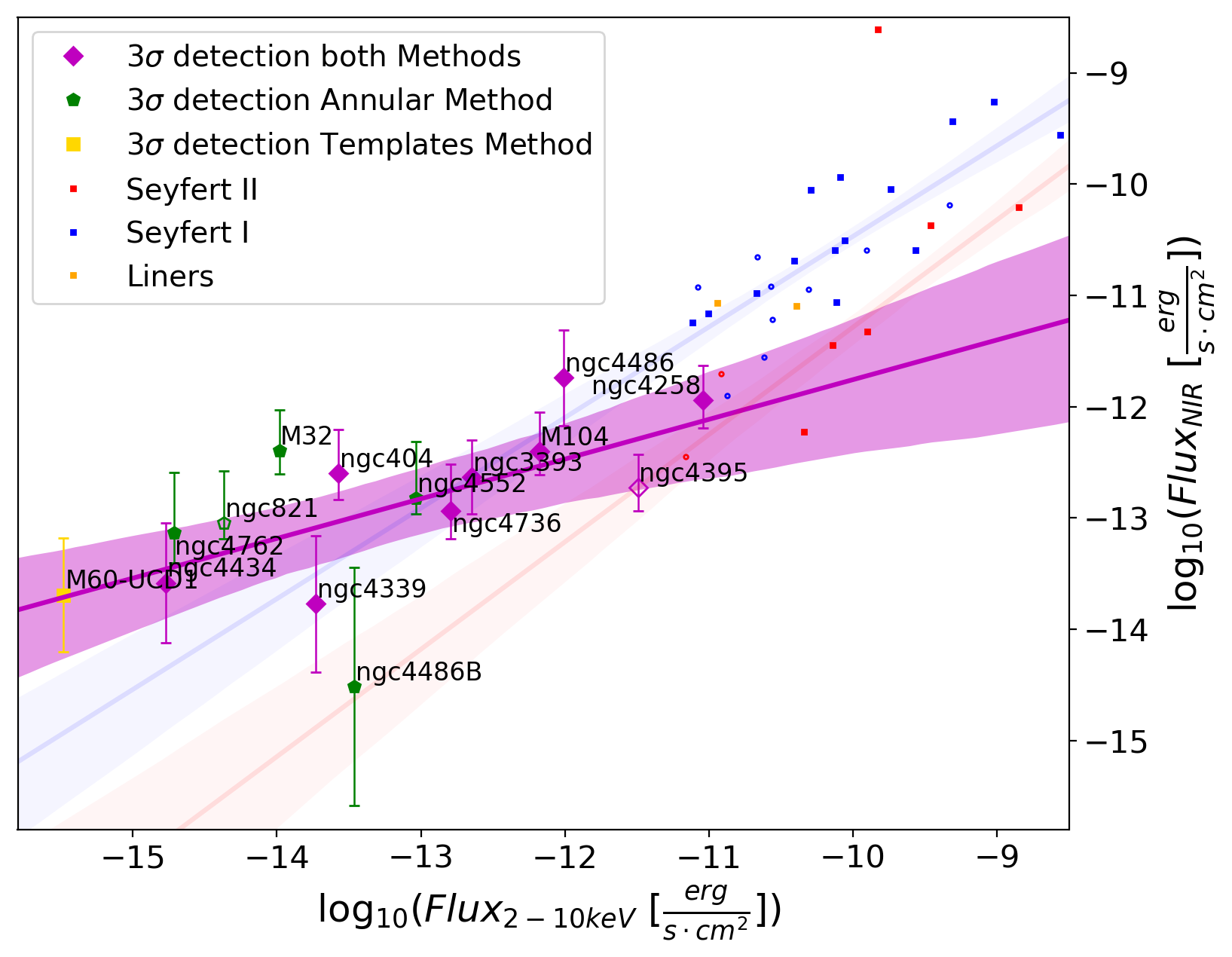}
\caption{The correlation between the NIR and 2-10 keV X-ray luminosities and fluxes.  The left panel shows the luminosity in the NIR (Table~~\ref{tab:table3}) versus the 2-10 keV X-ray luminosity (Table~\ref{tab:table1}). A best-fit regression to our data is shown in fuchsia. This regression includes AGNs of all types (both Types 1 and 2).  Small points show data from \citet{Muller-Sanchez2018} (squares) and \citet{Burtscher2015} (circles) for higher luminosity systems colored by AGN type.  Our sample is on average 3-4 orders of magnitude less luminous in the 2-10 keV X-ray and NIR than previous samples.  Separate fits to Seyfert I \& II galaxies by \citet{Burtscher2015} are shown as red and blue lines; these are both steeper than the relation we find for our lower-luminosity sample. \label{fig:NIRvsXray}
The right panel shows the correlation between our best NIR vs 2-10 keV X-ray fluxes. A correlation in fluxes provides stronger evidence for physical connection than a luminosity correlation, which can be created just by the distance dependence of the luminosity.}
\end{figure*}

Since the thermal AGN emission arises from reprocessing of more energetic photons, a correlation between the nuclear 2-10 keV X-ray nuclear and NIR fluxes is expected. This correlation has been observed in previous works for more luminous AGNs \citep{Quillen2001,ALONSO-HERRERO-2001,Burtscher2015,Muller-Sanchez2018}. Figure \ref{fig:NIRvsXray} shows the correlation for NIR and nuclear 2-10 keV X-ray luminosities and fluxes for our data -- we include all 15 galaxies, but code the data points according to the robustness of their detections, with magenta points being the most robust.
Small data points represent data from previous works (\citet{Burtscher2015} (circles) and \citet{Muller-Sanchez2018} (squares)) and are colored by AGN type.
The most notable feature of this plot is that our sample extends the luminosities down 3 orders of magnitude in 2-10 keV X-ray luminosity and 4 orders of magnitude in NIR luminosity. In addition our sample seems to follow a shallower correlation than more luminous AGN, suggesting a higher luminosity in the NIR relative to the previous correlation found by \citet{Burtscher2015}.  

We quantify the correlation in the $\log$ luminosities using linear regression and find:\\
\begin{equation}
\log{L_{nir,40}}=(0.50^{0.11}_{0.11})\cdot \log{L_{2-10 keV\:X-ray,40}}+0.49^{0.10}_{0.10}
\end{equation}\\

Where $L_{NIR,40}=\frac{L_{NIR}}{10^{40}\:(erg/s)}$ and $L_{2-10 keV\:X-ray,40}=\frac{L_{( 2-10 keV\:X-ray }}{10^{40}\:(erg/s)}$. The fluxes obtained from the spectra summed over the inner 0$\farcs$3 give a nearly identical slope (0.51). We also investigate the correlation in the fluxes in the the right panel in Figure~\ref{fig:NIRvsXray}. This provides stronger evidence of a physical link between NIR \& 2-10 keV X-ray than the luminosity correlation, since objects with uncorrelated fluxes could have correlated luminosities due to the distance dependence of the luminosities. To quantify the strength of the flux correlation we obtained a Spearman coefficient; the value of $r_{spearman}$=0.69, suggesting a $>$99\% significance of correlation (an $\alpha < 0.01$ for $N=15$). This estimate does not take into account the errors on the NIR fluxes; however if we bootstrap with resampled errors, the correlation remains significant  at the $\sim$97\% level. Combined with the NIR emission being confined to the nucleus, we regard this correlation as strong evidence that our observed NIR continuum emission has an AGN origin.  

\begin{figure*}[h!]
\plotone{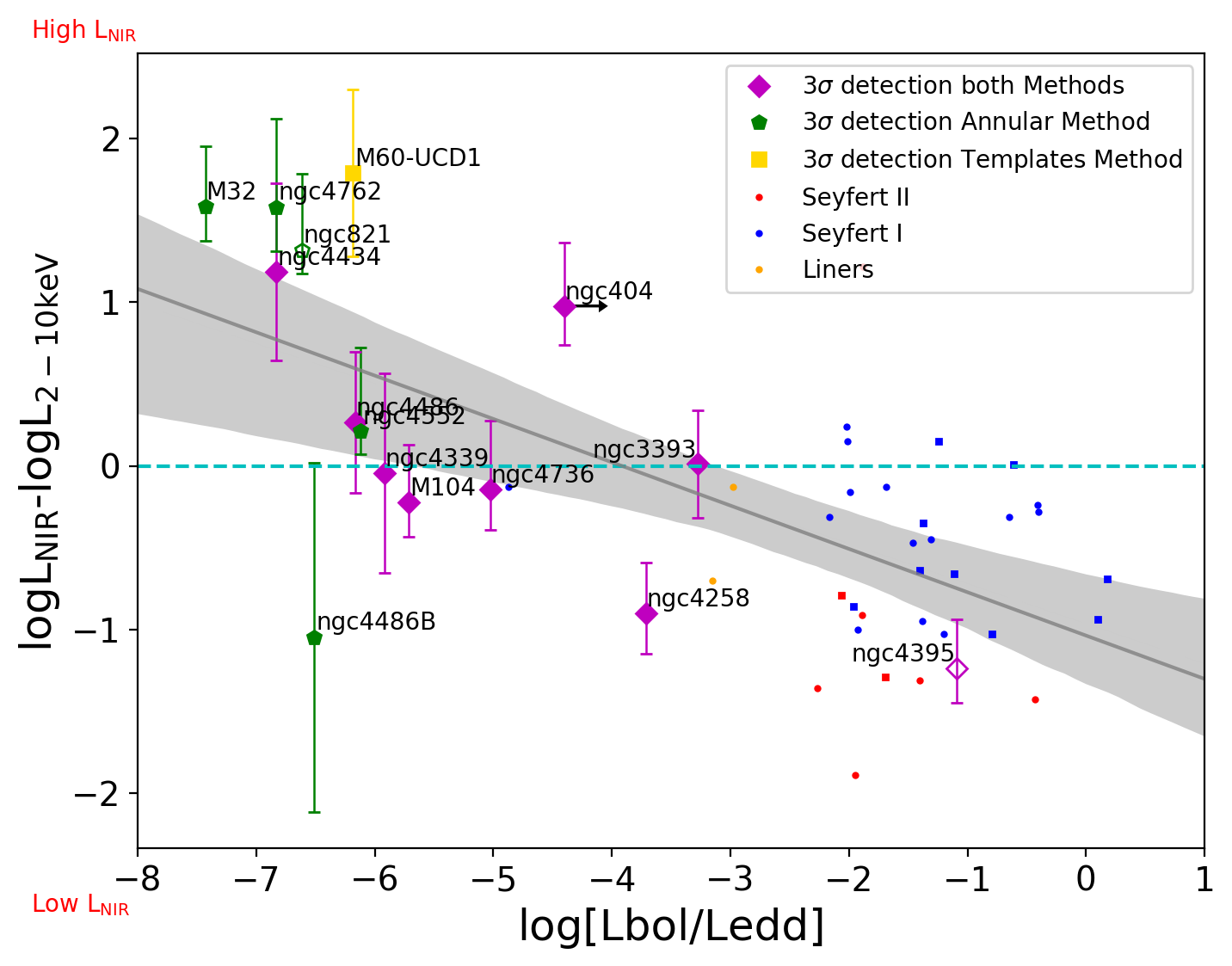}
\caption{The relation between NIR and 2-10 keV X-ray luminosities residual versus Eddington ratio. The bolometric luminosity was assumed to be $16\cdot L_{2-10 keV}$ \citet{Ho2008}. We obtained the Eddington's luminosity based on the dynamical BH mass listed in Table~\ref{tab:table1}, only an upper limit in the BH mass is available for NGC404. The small points show previous data from \citet{Burtscher2015} (circles)  \& \citet{Muller-Sanchez2018} (squares) and colored by AGN type; BH measurements from \cite{Saglia2016} were used to estimate their Eddington ratios. The best-fit regression for all the objects (our sample and previous measurements, without distinction on AGN type) is shown in grey; this suggest an excess of NIR emission at Eddington ratio $\leq -4$. \label{fig:bolumetric}}
\end{figure*}

The change in AGN SED with Eddington ratio may reveal changes in the physical structure of LLAGN.  Figure~\ref{fig:bolumetric} shows the relation of the residual of the NIR to 2-10 keV X-ray versus the Eddington ratio ($\log(L_{bol}/L_{edd})$). We estimated the bolometric luminosity of the AGN by assuming a bolometric correction to the (2-10 keV) luminosity of $16\cdot L_{2-10} keV$ \citet{Ho2008}. We note that this relation has a large scatter with an interquartile range of $9.6$. The Eddington luminosities were based on the dynamical BH masses listed in Table~\ref{tab:table1}. 

We find a clear result that the ratio of NIR to 2-10 keV X-ray luminosities increases at lower Eddington ratios.  To quantify this increase, we fit a line shown in grey for our sample as well as objects in the literature where BH mass estimates exist and found:\\
\begin{equation}
    \log(\frac{L_{NIR}}{L_{ 2-10 keV X-ray}})= -0.26^{0.05}_{0.05}\cdot (\frac{L_{bol}}{L_{Edd}}) -1.03^{0.08}_{0.08}
\end{equation}

In this case, the NIR luminosities from the spectra summed over the inner 0$\farcs$3 is a bit shallower ($-0.251$), but consistent within the errors. From this relation, we find that the NIR K-band luminosity outstrips the 2-10 keV X-ray luminosity at $L_{bol}/L_{edd}$ $\leq -4$. While we don't have a clear physical picture for the cause of this increase in NIR emission at lower Eddington ratios, there are a couple of plausible explanations:
\begin{itemize}
    \item The accretion mechanism and/or geometry could be changing, creating increased NIR reprocessing of disk emission (e.g.~a more spherical dust distribution).  To our knowledge, the expected infrared SED of a RIAF model has not yet been published in the literature. 
    \item The NIR emission could arise from a cold accretion disk (AD) instead of being produced by dust reprocessing. \citet{Laor2011} found that the temperature of the accretion disk decreases at lower Eddington ratios, and that objects with $L_{bol}/L_{edd}$ $\leq -6$ have disk temperatures of $\sim\:3000\:K$, therefore the peak emission for the AD will be in the NIR instead of the UV. For the AD to be the mechanism powering the LLAGNs, the inner radius of the AD needs to be a few tens of the gravitational radius ($r_{g}=GM/c^{2}$) to produce a significant flux.  This cold accretion disk scenario was at least partially confirmed in one LLAGN recently  \citep{Bianchi2019}.
    \item Jet emission could start contributing increasingly to the NIR emission at these low accretion rates.  Models showing strong jet emission in NGC4486/M87 have been published by \citet{Prieto2016}, and in general, jet emission is expected to be increasingly important for LLAGN \citep[e.g.][]{Ho2008, Trump2011}.
\end{itemize}

Regardless of the origin of this emission, the large NIR luminosity we observe for LLAGN suggests the NIR could serve as a more sensitive AGN indicator than X-rays. This suggests that future observations with JWST will reveal many faint AGN at the centers of galaxies across cosmic history.  Understanding the spectral signature of this emission will still be required to separate it from stellar emission \citep[e.g.][]{Kirkpatrick2015,Satyapal2018}.

\section{Alternative to AGN origin \label{sec:non-agn}}
The evidence that the NIR continuum emission we observe has an AGN origin comes primarily from the correlation of that emission with the 2-10 keV X-ray flux, which is a good tracer of AGN activity even to low luminosities \citep[e.g.][]{Miller2012,Gallo2018}. Additionally, the NIR sources are compact and located at the centers of each of the objects. No similar high flux areas are seen off center. These sources are also associated with weaker CO line strengths in most objects, and a redder continuum as expected for weaker versions of the NIR AGN flux that we see at higher luminosities (see Figure \ref{fig:CO_continuum_plot}).

Despite this, we discuss other sources of NIR continuum and hot dust emission, as these sources may serve a contaminating role, especially at lower luminosities. Obscuration from dust could produce the reddening of the slope that we see in left panel of Figure~\ref{fig:maps}, where bluer wavelengths of the spectra will be more extincted than redder wavelengths. An estimation of the necessary $A_{V}$ to produce the reddening was estimated using the infrared-optical extinction relation described in \citet{Cardelli1989}, the result are shown in Table~\ref{tab:table3}.  In general most of these values are reasonable given the $A_{V} \sim 30$ mags seen in the Milky Way. However, dust extinction cannot explain the weakening of the CO linestrengths and the concentration towards the center, and thus dust absorption is not a plausible explanation for our observations.  

Another alternative scenario would be the ubiquitous presence of enshrouded AGB stars (often called extreme AGB stars) in the central 0$\farcs$3 of our sample galaxies. Normal AGB stars have strong CO bandheads, and thus can't be causing the signature we see.  However, extreme AGB stars can undergo significant mass loss, creating a circumstellar envelope that obscures the photosphere.  These stars thus have the bolometric luminosity of a bright AGB star but the temperature of the circumnuclear dust around them, and lack strong CO bandheads \citep[e.g.][]{Blum2014}, and thus match the signature of the continuum emission we find at the centers of our galaxies. However, these extreme AGB stars represent only a small fraction of the AGB population \citep{Blum2006}, and a very small fraction of the K-band flux \citep[$<$2\% of the K-band flux in the Magellanic Clouds;][]{Melbourne2013}

We test the possibility of extreme AGB contamination in M32, our faintest galaxy ($L_{NIR} = 2.51\cdot10^{37} \frac{erg}{s}$), and one with an available dynamical mass measurement \citep{Nguyen2018}.  Given the high stellar mass of this nucleus and the faint emission, this is the most likely object for extreme AGB contamination.  Using the PARSEC-COLIBRI isochrones \citep{PARSEC-COLIBRI,Pastorelli2019}, we have tried to estimate the minimum mass required to mimic the hot dust signature we see.  To be conservative and maximize the AGB luminosity, we use a 1 Gyr population and cut at J-K$_s > 2$ to select extreme AGB stars \citep{Melbourne2013,Pastorelli2019}.  We find the total K-band luminosity in extreme AGB stars per unit mass of $2.0\cdot10^{31}\:\frac{erg}{s\:M_{\odot}}$.  Using this, the minimum mass required to mimic this signature is therefore $1.3\cdot10^{6}\:M_{\odot}$. 

We can compare this to the dynamical mass of M32. The nucleus in M32 shows no significant gradient in color with radius \citep{Lauer1998}, suggesting a constant stellar population and therefore $M/L$.  We can therefore estimate the dynamical mass in our 0$\farcs$3 radius aperture by noting that it contains $\sim$10\% of the total light of the nuclear star cluster, and using the dynamical mass estimate of this cluster by \citep{Nguyen2018} to obtain a total mass in this aperture of $\sim 1.7\cdot10^{6}\:M_{\odot}$.  So it appears feasible to explain the NIR continuum in M32 with an extreme stellar population.  However, we still prefer an AGN interpretation for the continuum emission for two reasons: (1) spectral synthesis analysis of the ages of stars in the nucleus suggest a typical age of 4~Gyr with small fractions of the mass at younger ages \citep{Coelho2009}, which would lead to a much lower XAGB luminosity fraction, and (2) there is no color gradient suggesting a stellar population variation within the nucleus of M32 \citep{Lauer1998}, while the hot dust signature is clearly confined to the central region. 


Overall, we prefer the AGN interpretation for the continuum emission seen in our target galaxies given its compactness and ubiquity, and the correlation in the flux of this emission with the X-ray flux.

\section{Discussion \& Conclusions\label{sec:conclusion}}

In this paper, we used Gemini/NIFS IFU adaptive optics data to examine the NIR emission in 15 LLAGNs. Most of the galaxies show a clear reddening of the continuum near the center of the galaxy.  Assuming the source of this emission is the same hot dust emission previously seen in higher luminosity AGN, we decompose the integral field data cubes into stellar and hot dust components.  We find this thermal hot dust emission is concentrated at the center, and confirm that the CO bandhead absorption in these regions are weaker in a majority of the objects.  Furthermore, this thermal NIR emission is correlated with the 2-10 keV X-ray in the sources with a Spearman coefficient of $r_{s}=0.69$ indicating a $>$99\% significance of correlation.  

This work is the first to systematically look at the NIR emission from LLAGN.  The mean 2-10 keV X-ray luminosity of our sample is 38.8 ergs/s, $\sim$4 orders of magnitude below previous similar studies of the NIR emission in AGN  \citep{Burtscher2015,Muller-Sanchez2018}.  
At these lower luminosities, we find a shallower correlation between the NIR and 2-10 keV X-ray luminosities.  Furthermore, the ratio of the NIR to 2-10 keV X-ray flux appears to increase with decreasing Eddington ratio, suggesting a potential change in the physical mechanism creating the NIR emission at these lowest luminosities. There is a possibility that some of the X-ray detections, especially at the lowest luminosity end, are due to non-AGN emission (i.e. from X-ray binaries) -- in this case, the ratio of NIR to 2-10 keV X-ray luminosities would be even higher than we find here.
These results are somewhat puzzling, considering theoretical studies predict the torus will disappear at low-luminosities \citep[e.g.][]{elitzur2006,Honig2007}.

The bright NIR emission seen in the lowest luminosity AGN suggests that observations at these wavelengths can be a useful tool for detecting LLAGN. The IR has been proven to be a reliable AGN selection method using different color filters at $3.4\mu$m, 4.6$\mu$m,12$\mu $m, 22$\mu$m ($W1 ,W2 ,W3 ,W4$ resp.) in the Wide-field Infrared Survey Explorer (WISE) \citep{Jarrett2011,Mateos2012,Stern2012,Hainline2016}.  However, we find that typically, our sample is not detectable at the $\sim6\arcsec$ resolution of WISE.  Although we cannot easily evaluate the galaxy luminosity at this resolution due to NIFS $3\arcsec$  field-of-view, we can get a lower limit on the galaxy contribution of our AGN sources on this scale by comparing our data on the largest scale available.  We, therefore, compared the host galaxy flux within an aperture of 1$\farcs$5 radius with the best-fit hot dust flux from Table~\ref{tab:table3} at $3.4\mu$m and $4.6\mu$m. We assume the host galaxy is a pure blackbody with a stellar temperature constrained by the spectrum in the $K$-band (typically $\sim$3500K), and the hot dust emission is a blackbody with the best-fit temperature from Table~\ref{tab:table3}. We find that the median AGN contribution in a 1$\farcs$5 aperture is 20 \& 9 times less luminous than the host galaxies stellar light at $3.4\mu$m and $4.6\mu$m respectively.  The only case where the AGN was brighter in a 1$\farcs$5 aperture was in NGC4395 where the AGN 
component is calculated to be $1.5$ \& $3$ times brighter than the host galaxy at $3.4\mu$m and $4.6\mu$m  respectively.

Looking towards the future, JWST imaging and spectroscopy will provide unparalleled sensitivity at wavelengths beyond 2$\mu$m, and will have comparable resolution to the adaptive optics data we have presented here.  For an aperture of radius 0$\farcs$3 and extrapolating the dust and stellar components based with their blackbody curves to longer wavelengths as above, we expect the AGN light to be dominant at wavelengths higher than $4\mu$m for a majority of our galaxies.  Even when the light from the AGN doesn't dominate the SED, we expect techniques similar to those used here can be used to separate AGN and stellar emission, especially with IFU observations using NIRSpec and MIRI, enabling reaching even fainter sources than those detected here.  Additional observations extending the wavelength coverage of this NIR AGN emission with JWST should reveal the physical nature of the NIR AGN emission in these lowest luminosity systems.  

\acknowledgments

We thank the anonymous referee for their valuable comments and Martha Boyer and Robert Blum for informative discussions on X-AGB stars.  AD and ACS acknowledge support from NSF AST-1350389. JS acknowledges support from NSF grant AST-1514763 and a Packard Fellowship.  Based on observations obtained at the Gemini Observatory and acquired through the Gemini Observatory Archive, which is operated by the Association of Universities for Research in Astronomy, Inc., under a cooperative agreement with the NSF on behalf of the Gemini partnership: the National Science Foundation (United States), National Research Council (Canada), CONICYT (Chile), Ministerio de Ciencia, Tecnolog\'{i}a e Innovaci\'{o}n Productiva (Argentina), Minist\'{e}rio da Ci\^{e}ncia, Tecnologia e Inova\c{c}\~{a}o (Brazil), and Korea Astronomy and Space Science Institute (Republic of Korea).  
%

\vspace{5mm}
\facilities{Gemini/NIFS}



\FloatBarrier

\appendix
Here we show the hot dust maps and the CO bandhead absorption line depth against the fitted slope of the continuum between ($2.05-2.28\:\mu$m) for the rest of the sample.
\FloatBarrier
\begin{figure*}
\gridline{\fig{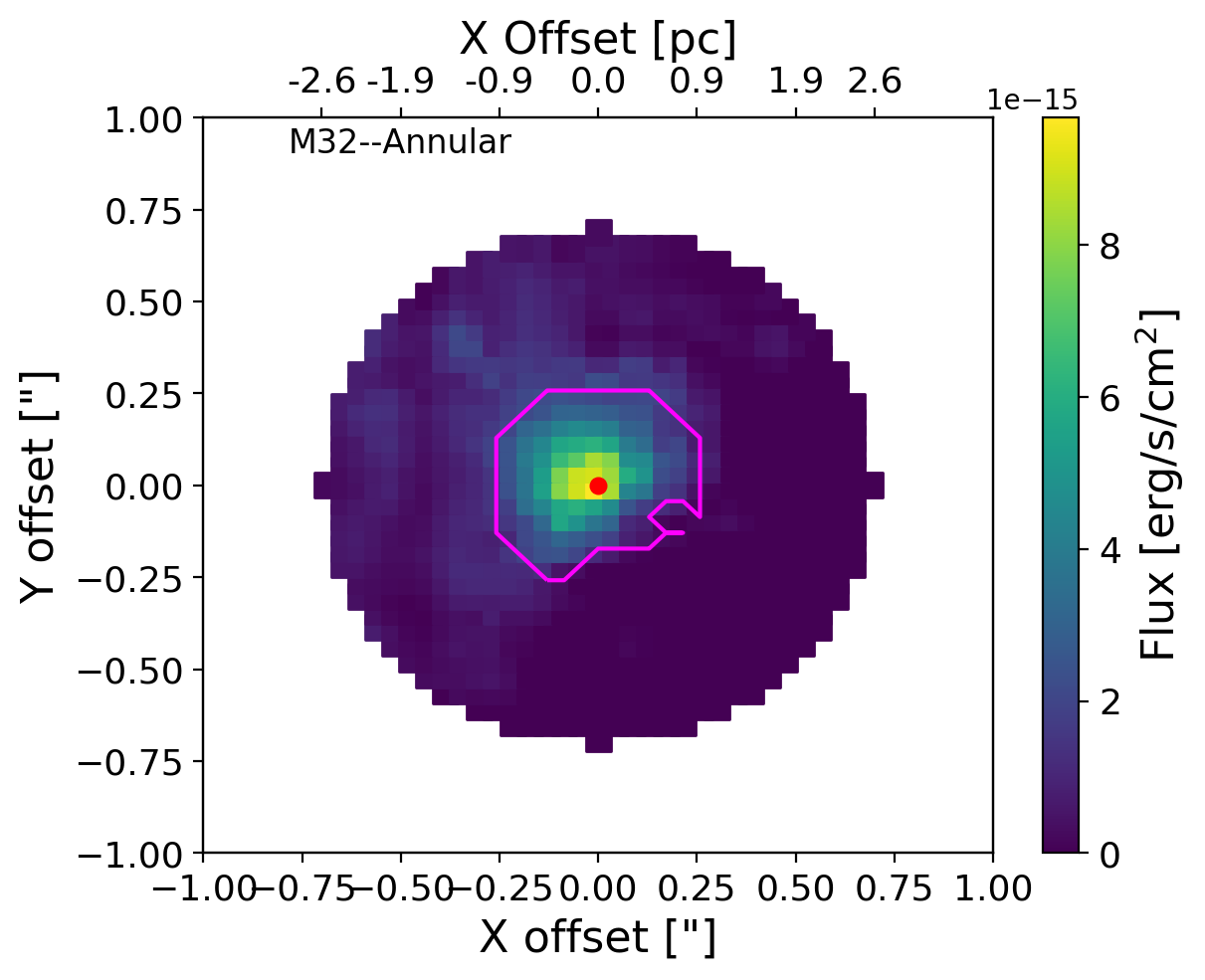}{0.5\textwidth}{(a)}
          \fig{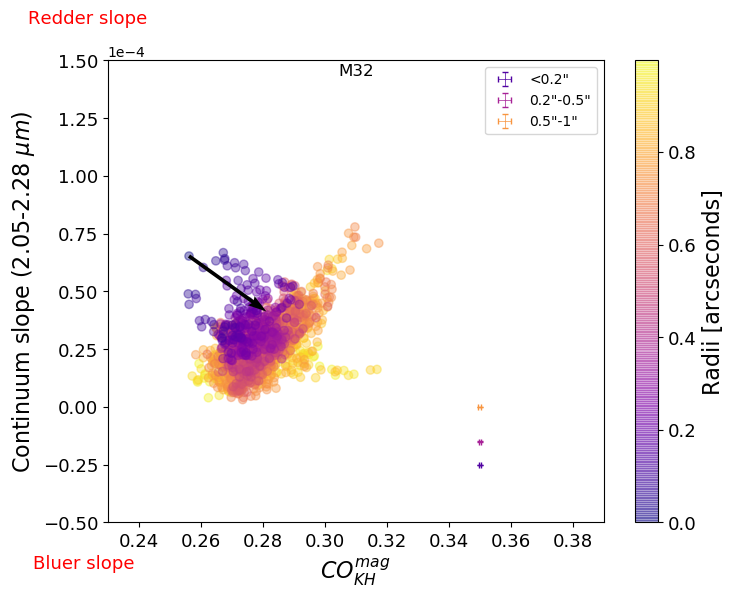}{0.5\textwidth}{(b)}}
\gridline{\fig{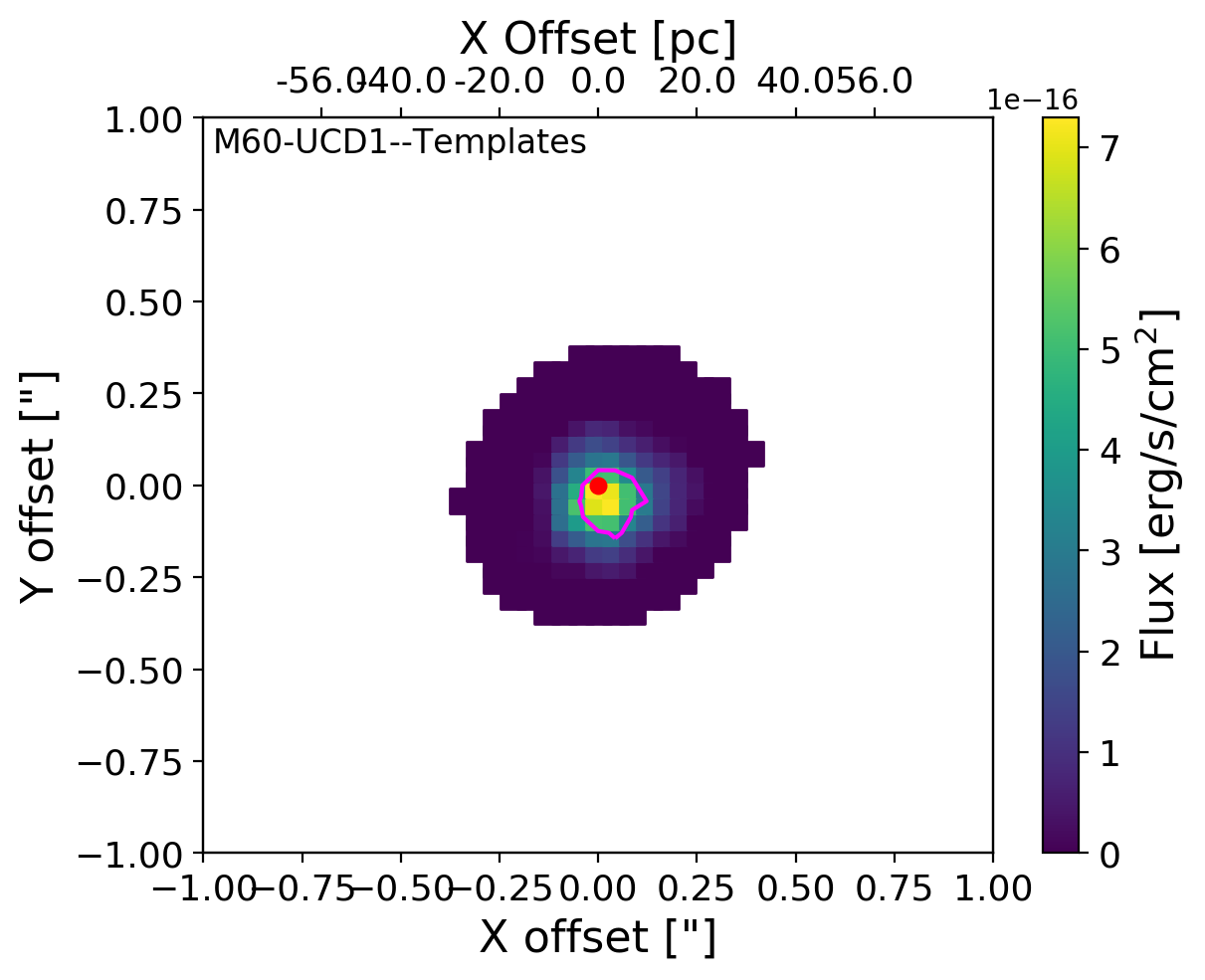}{0.5\textwidth}{(a)}
          \fig{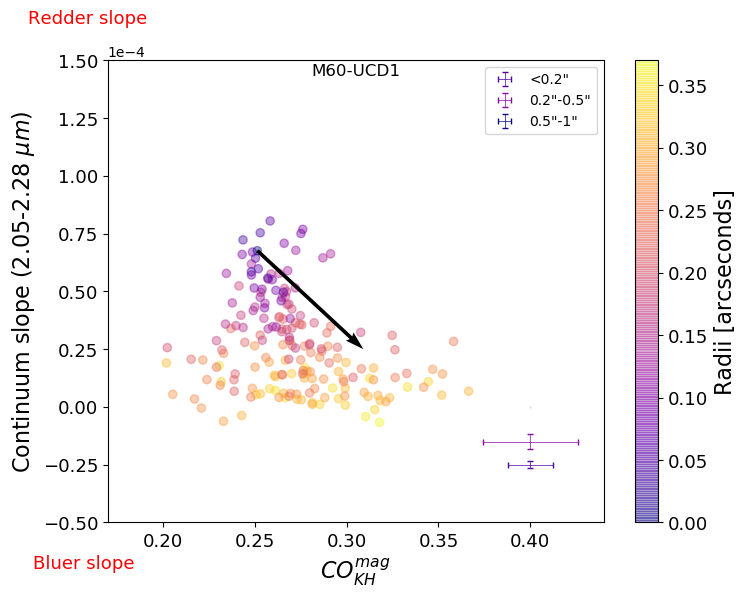}{0.5\textwidth}{(b)}}

\end{figure*}

\begin{figure*}
\gridline{\fig{ngc404.png}{0.5\textwidth}{(a)}
          \fig{ngc404_CO_ew.png}{0.5\textwidth}{(b)}}
\gridline{\fig{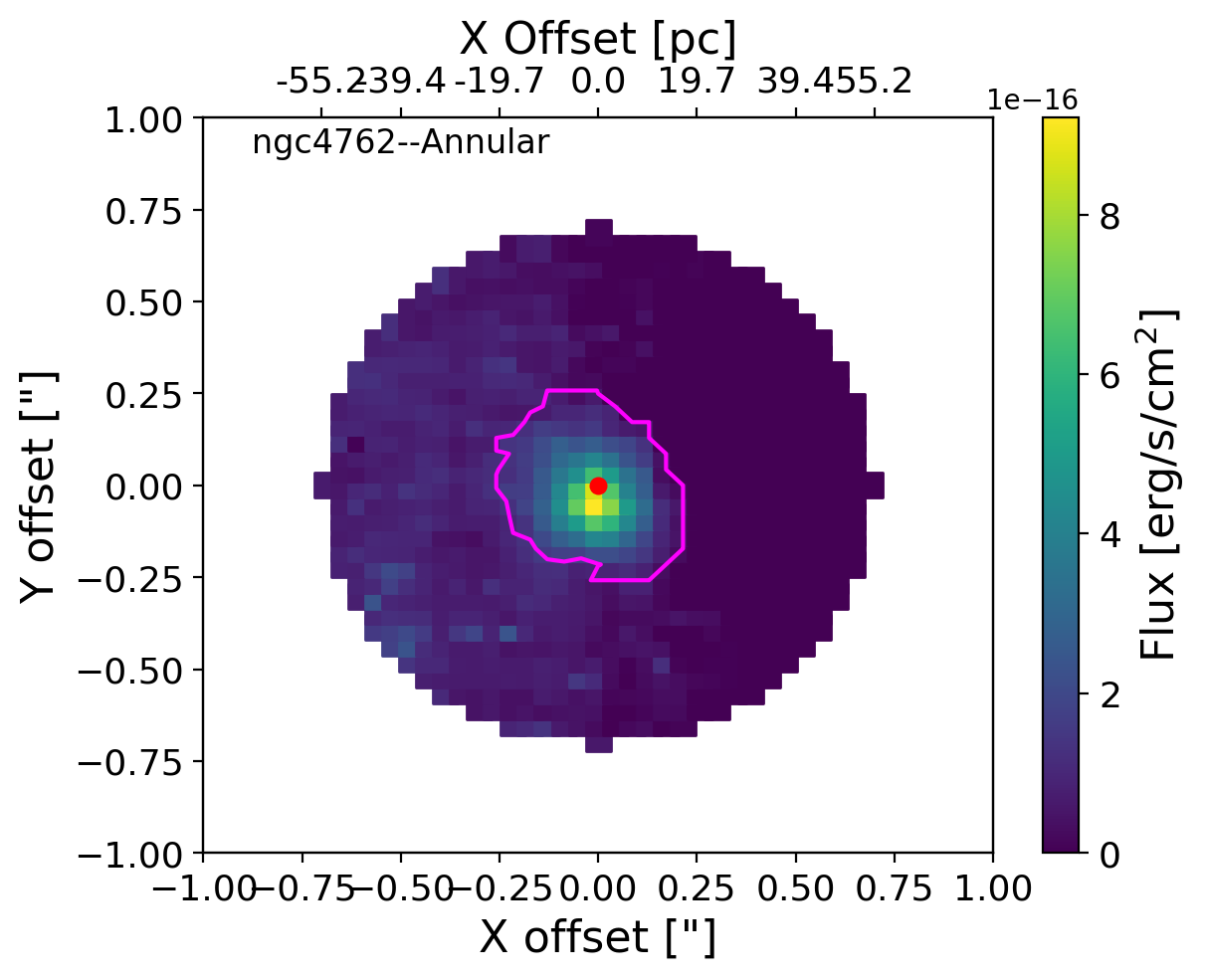}{0.5\textwidth}{(a)}
          \fig{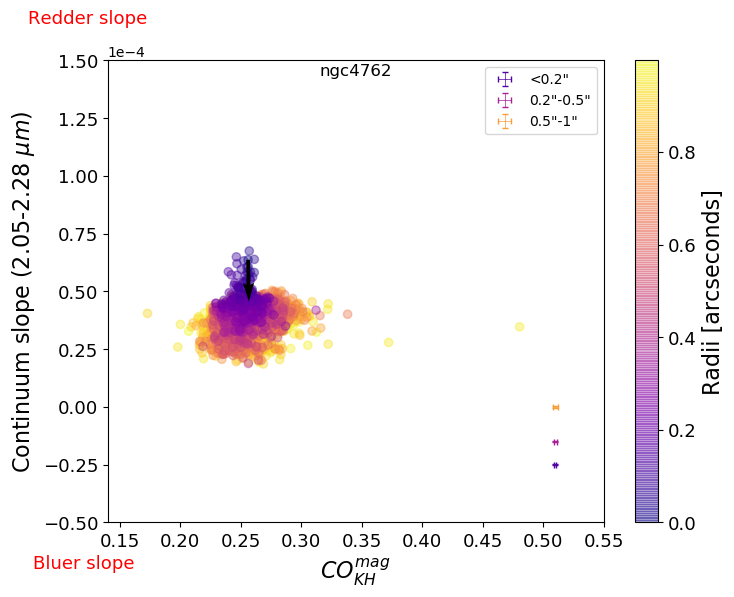}{0.5\textwidth}{(b)}}
\gridline{\fig{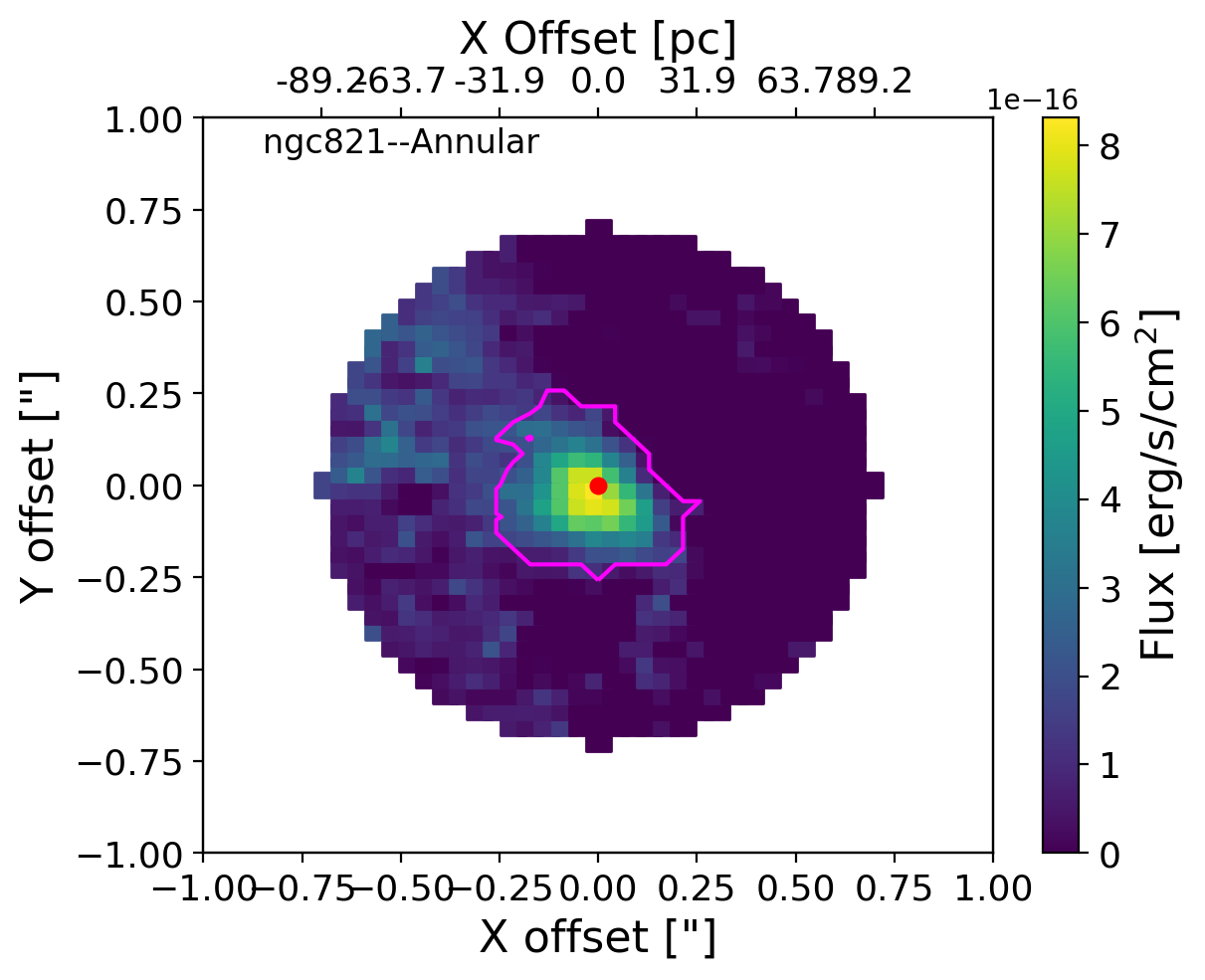}{0.5\textwidth}{(a)}
          \fig{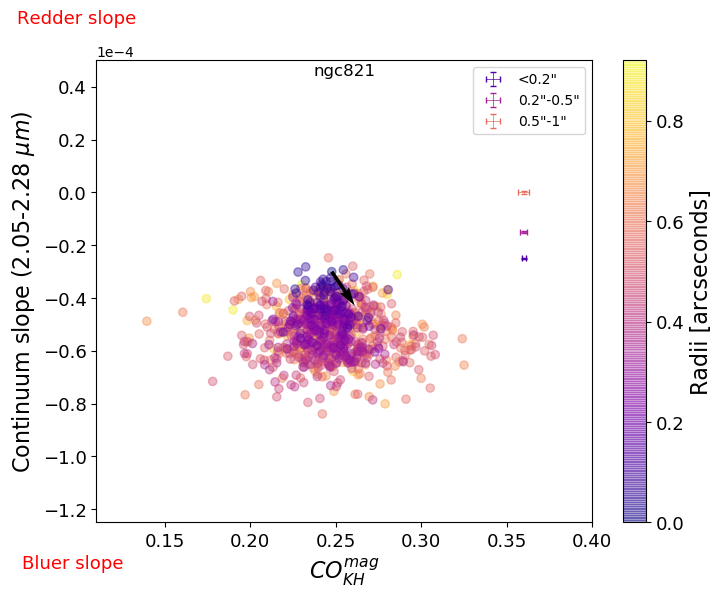}{0.5\textwidth}{(b)}}
\end{figure*}

\begin{figure*}[h!]
\gridline{\fig{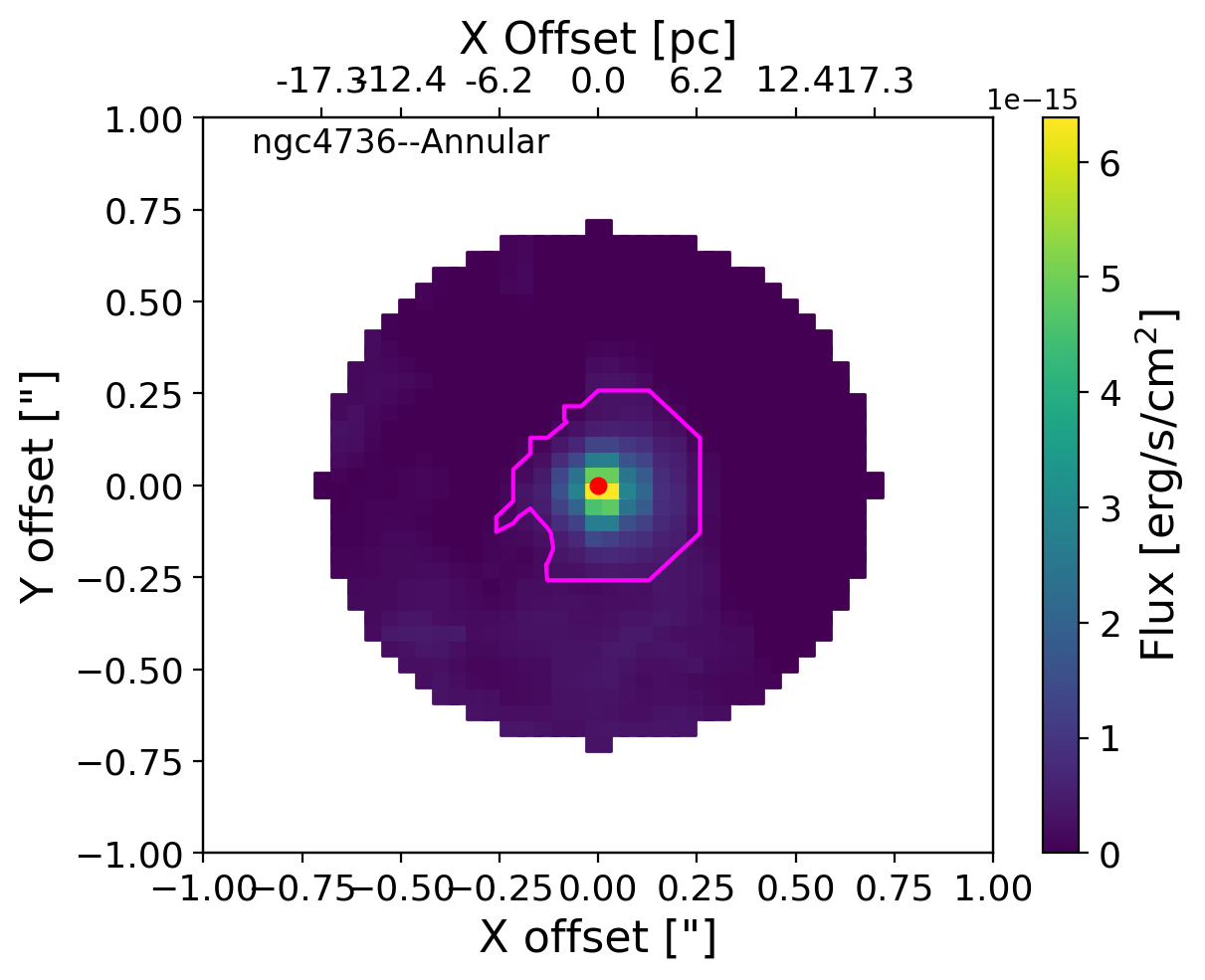}{0.5\textwidth}{(a)}
          \fig{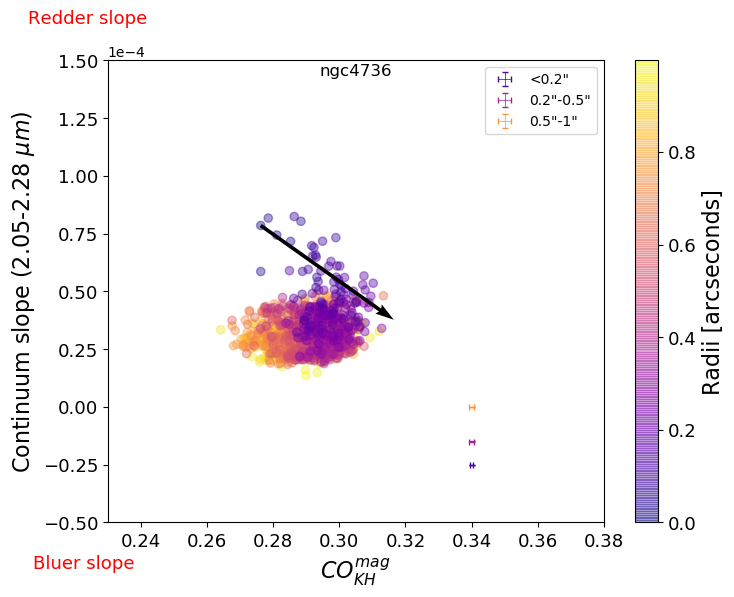}{0.5\textwidth}{(b)}}
\gridline{\fig{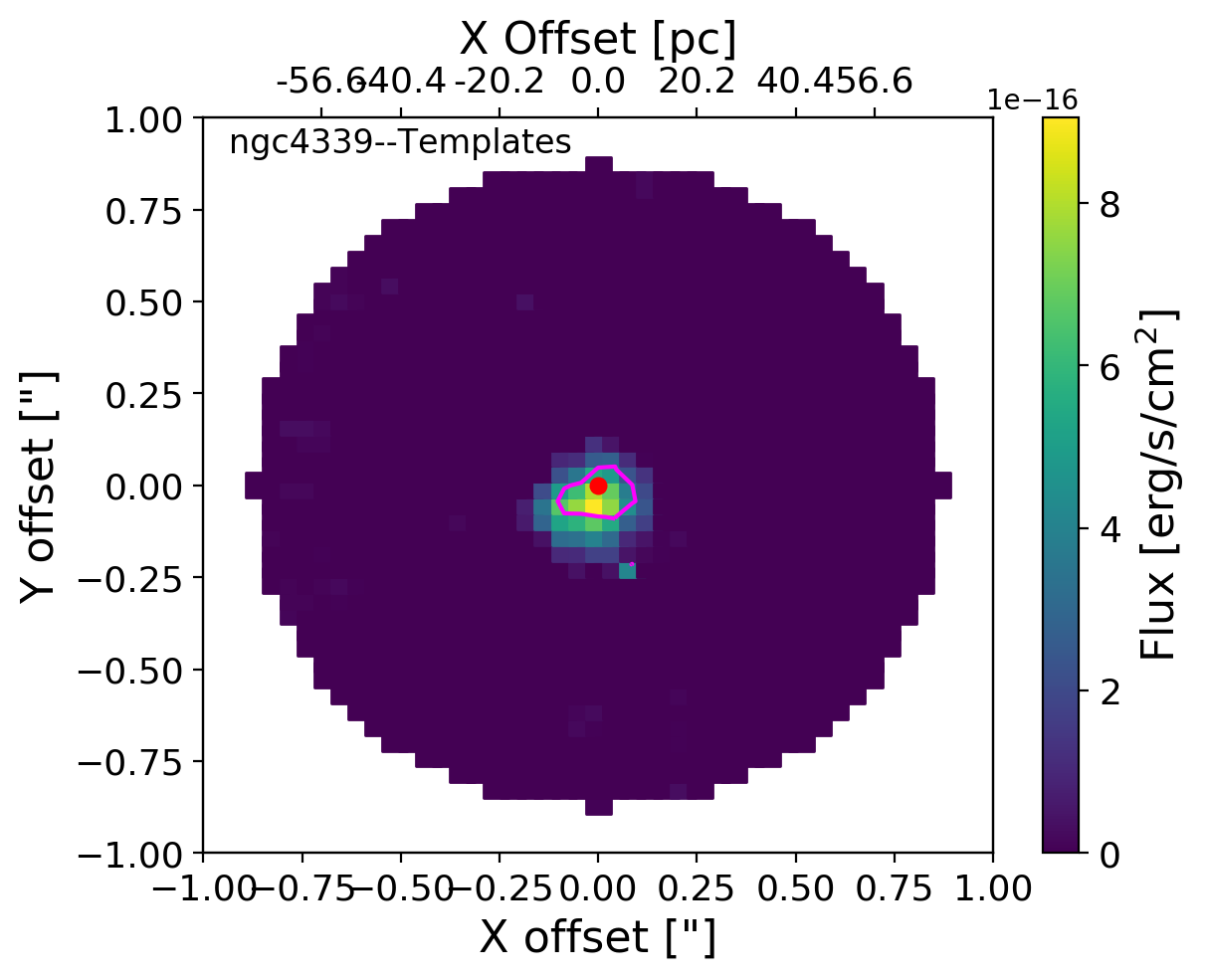}{0.5\textwidth}{(a)}
          \fig{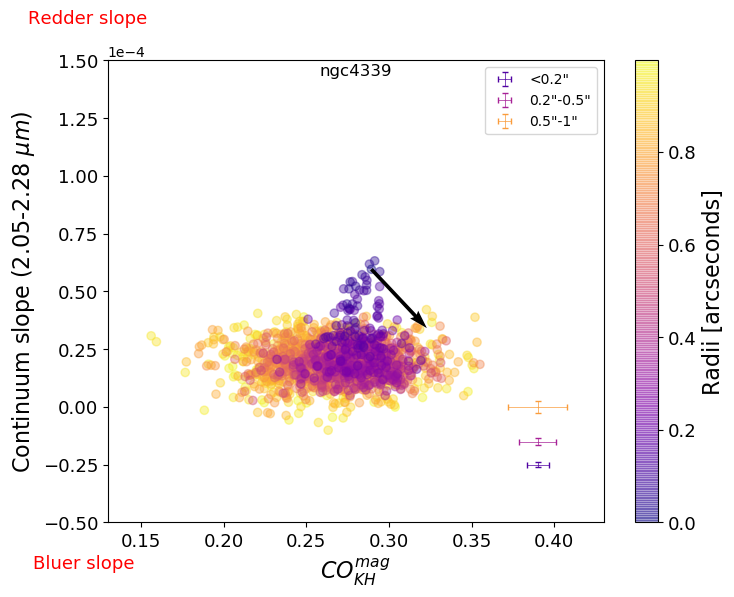}{0.5\textwidth}{(b)}}
\gridline{\fig{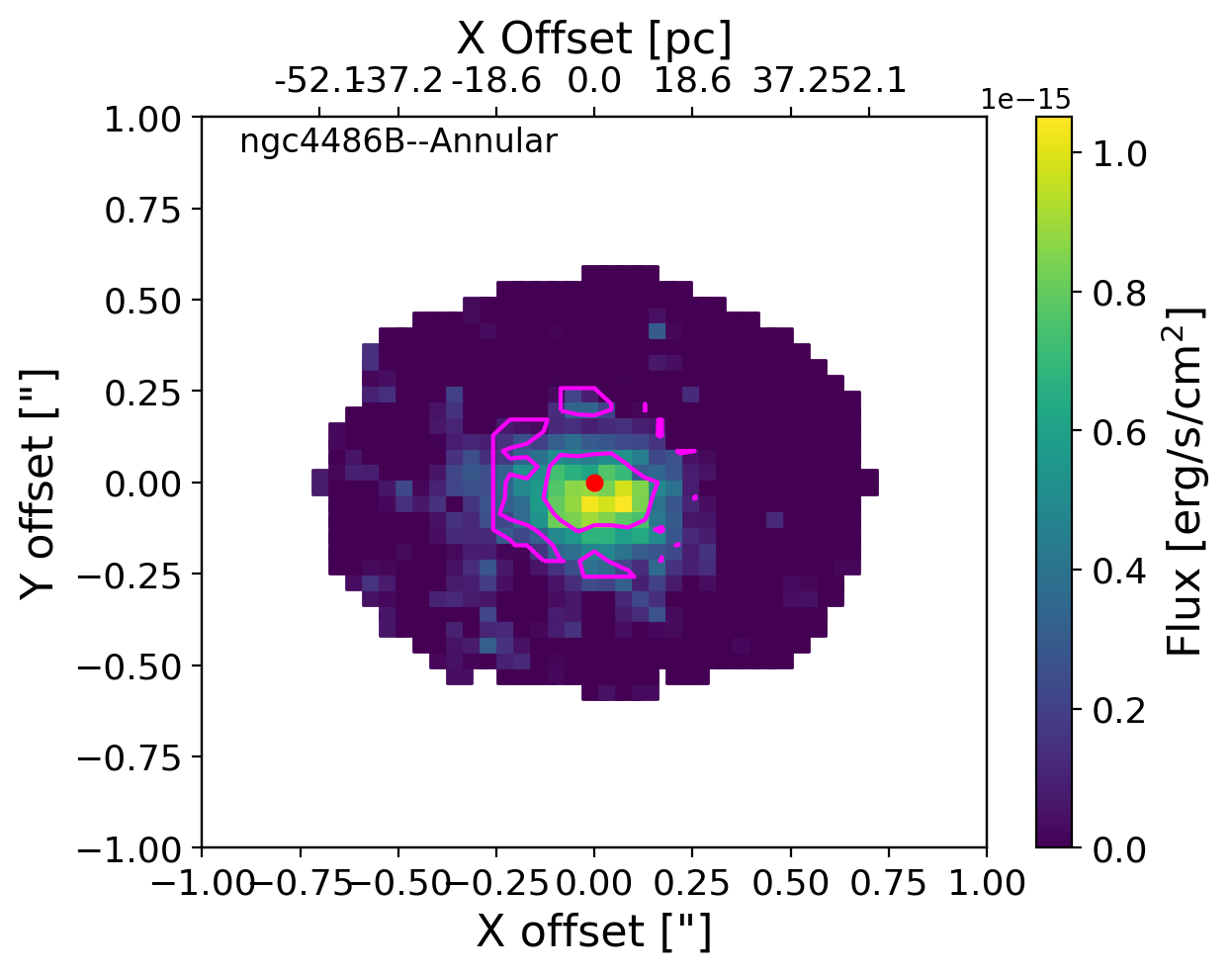}{0.5\textwidth}{(a)}
          \fig{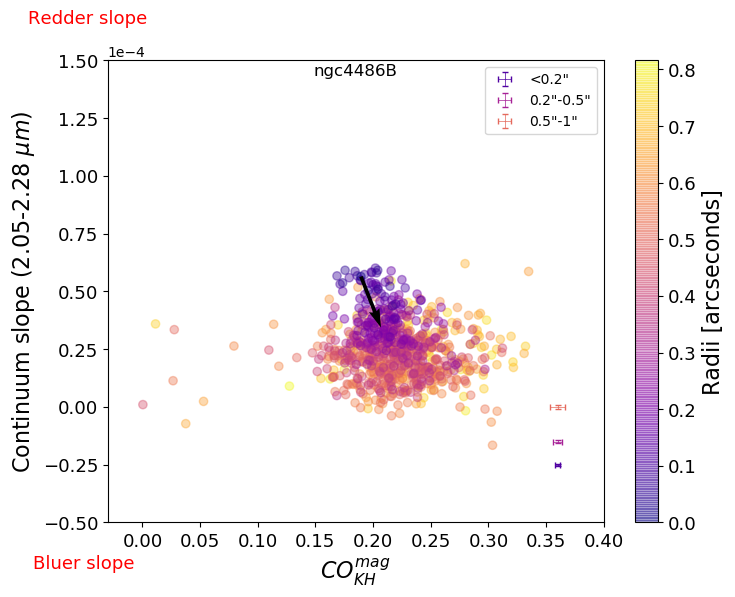}{0.5\textwidth}{(b)}}
\end{figure*}
\begin{figure*}[h!]
\gridline{\fig{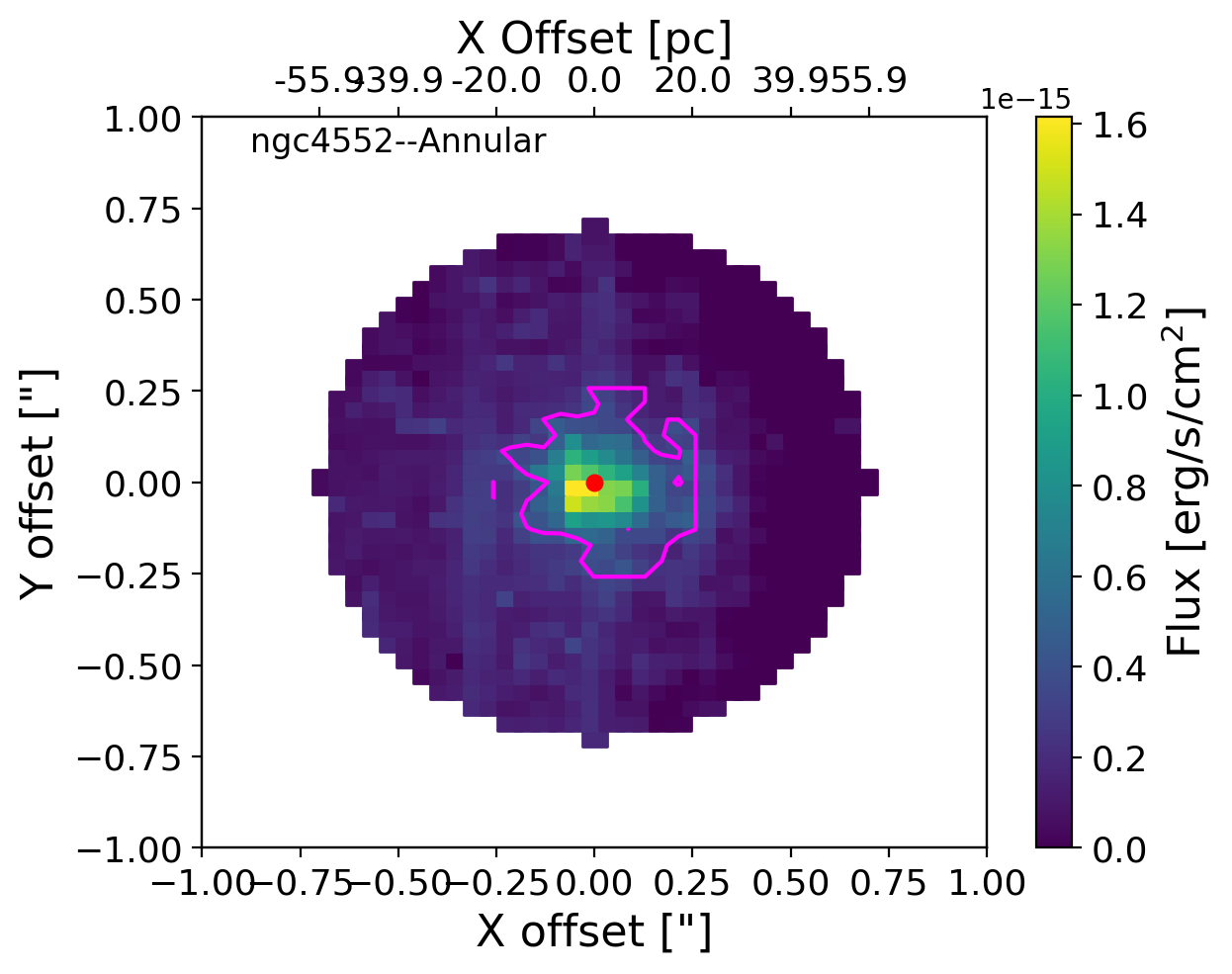}{0.5\textwidth}{(a)}
          \fig{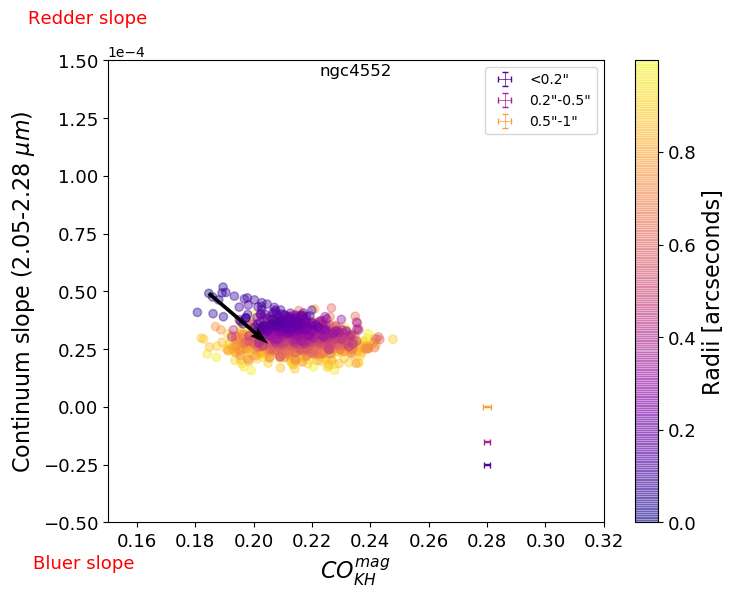}{0.5\textwidth}{(b)}}
\gridline{\fig{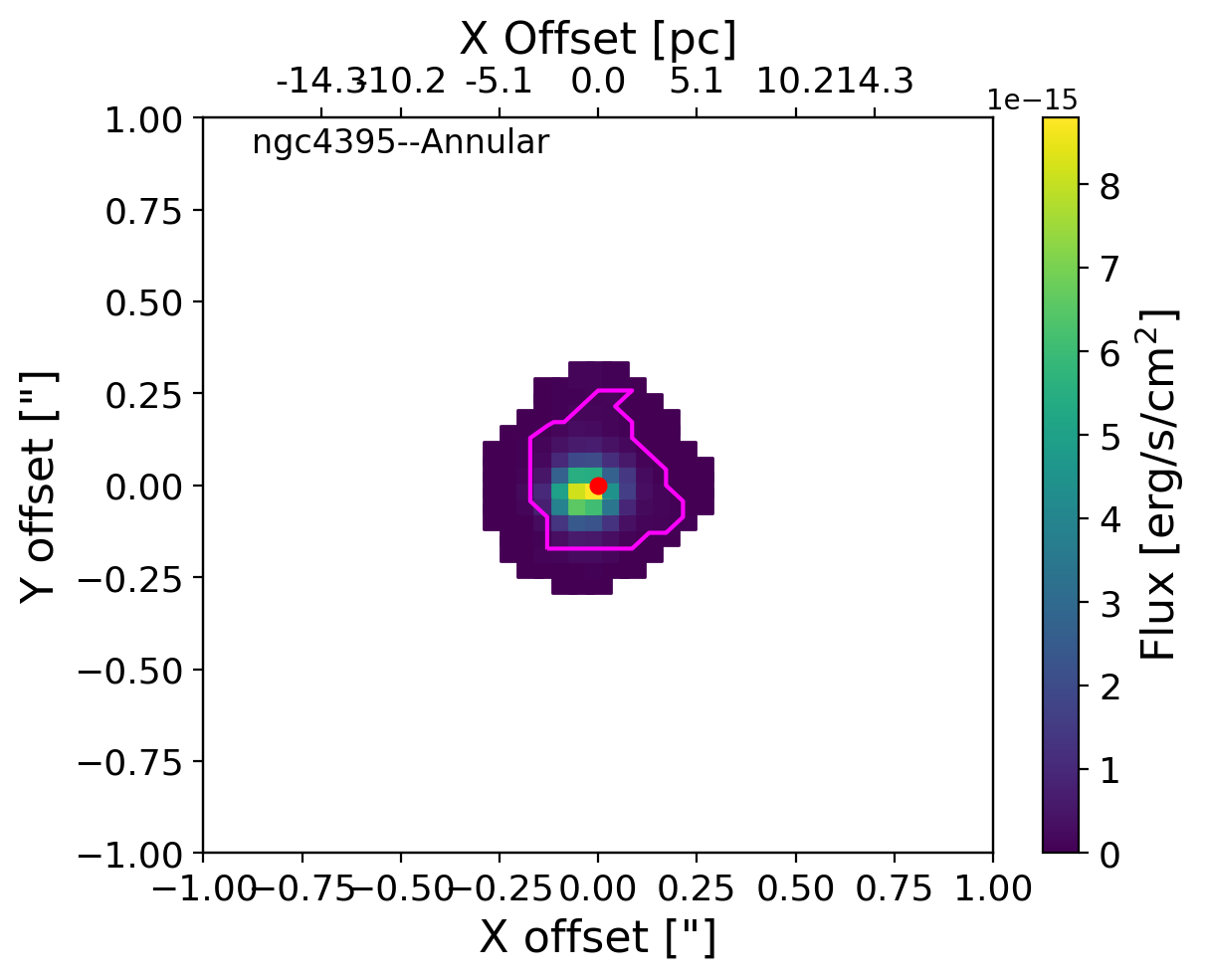}{0.5\textwidth}{(a)}
          \fig{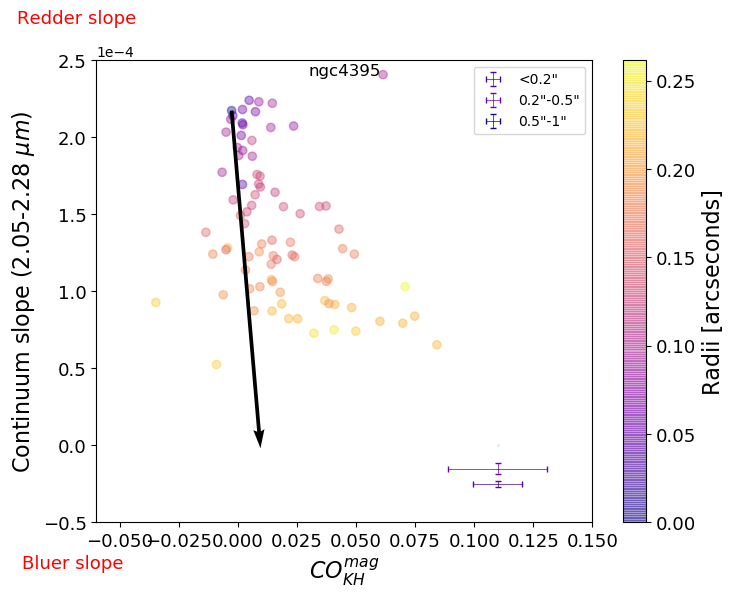}{0.5\textwidth}{(b)}}
\gridline{\fig{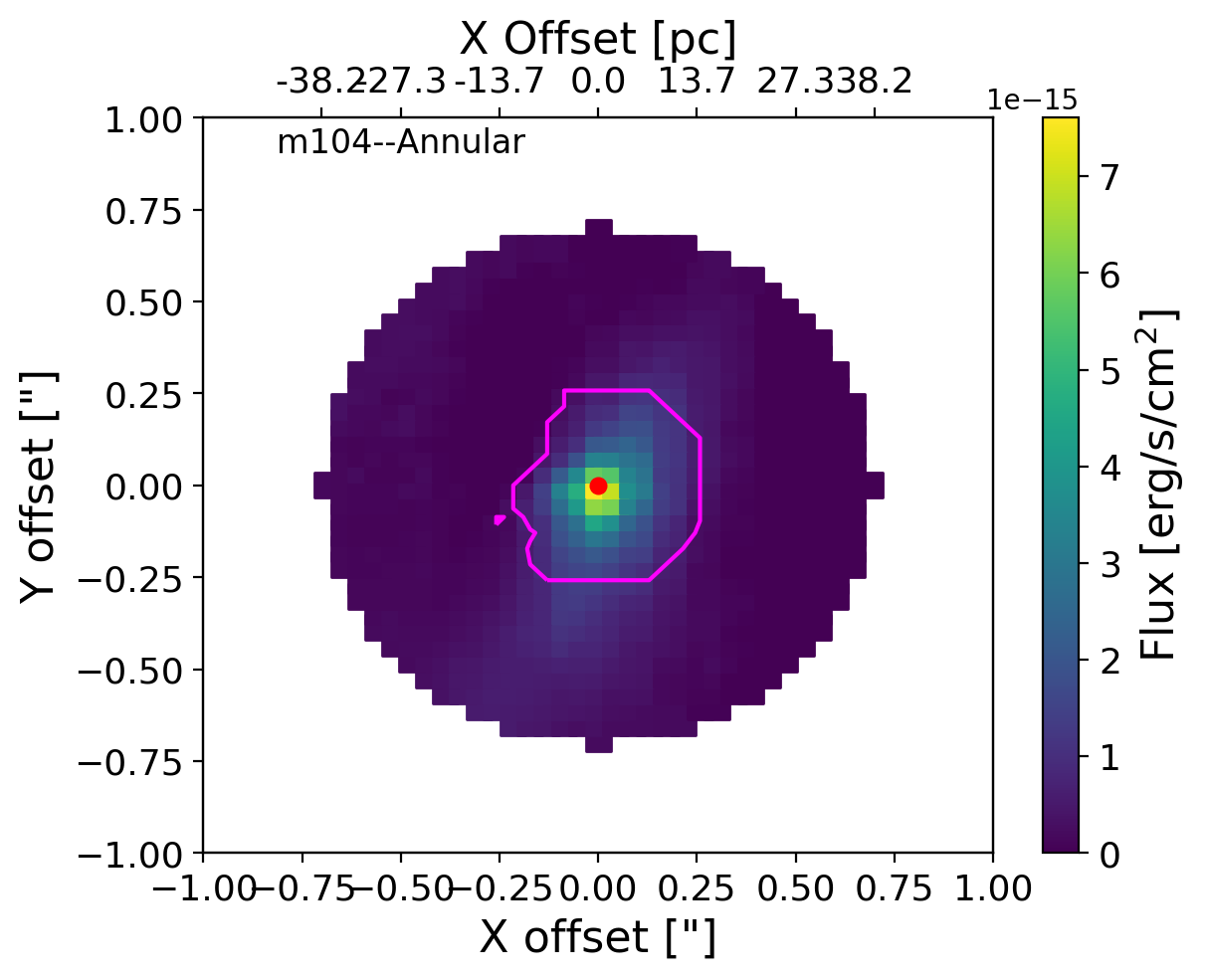}{0.5\textwidth}{(a)}
          \fig{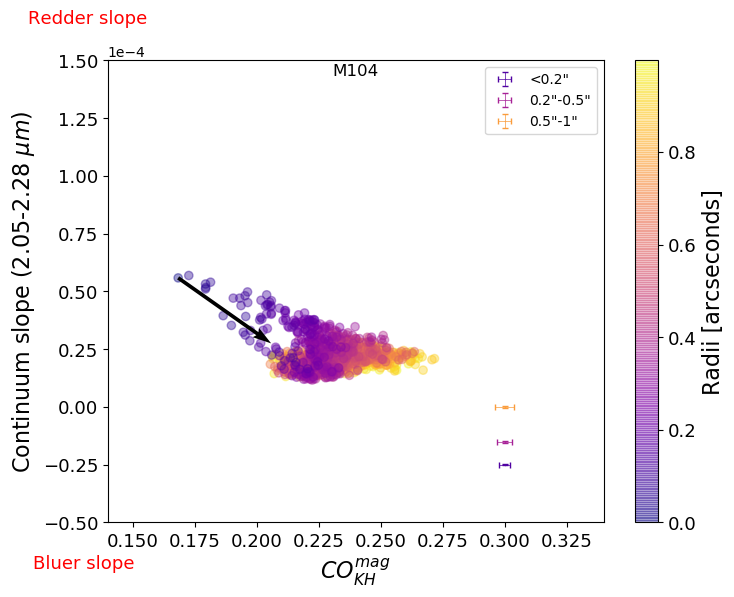}{0.5\textwidth}{(b)}}
\end{figure*}
\begin{figure*}
\gridline{\fig{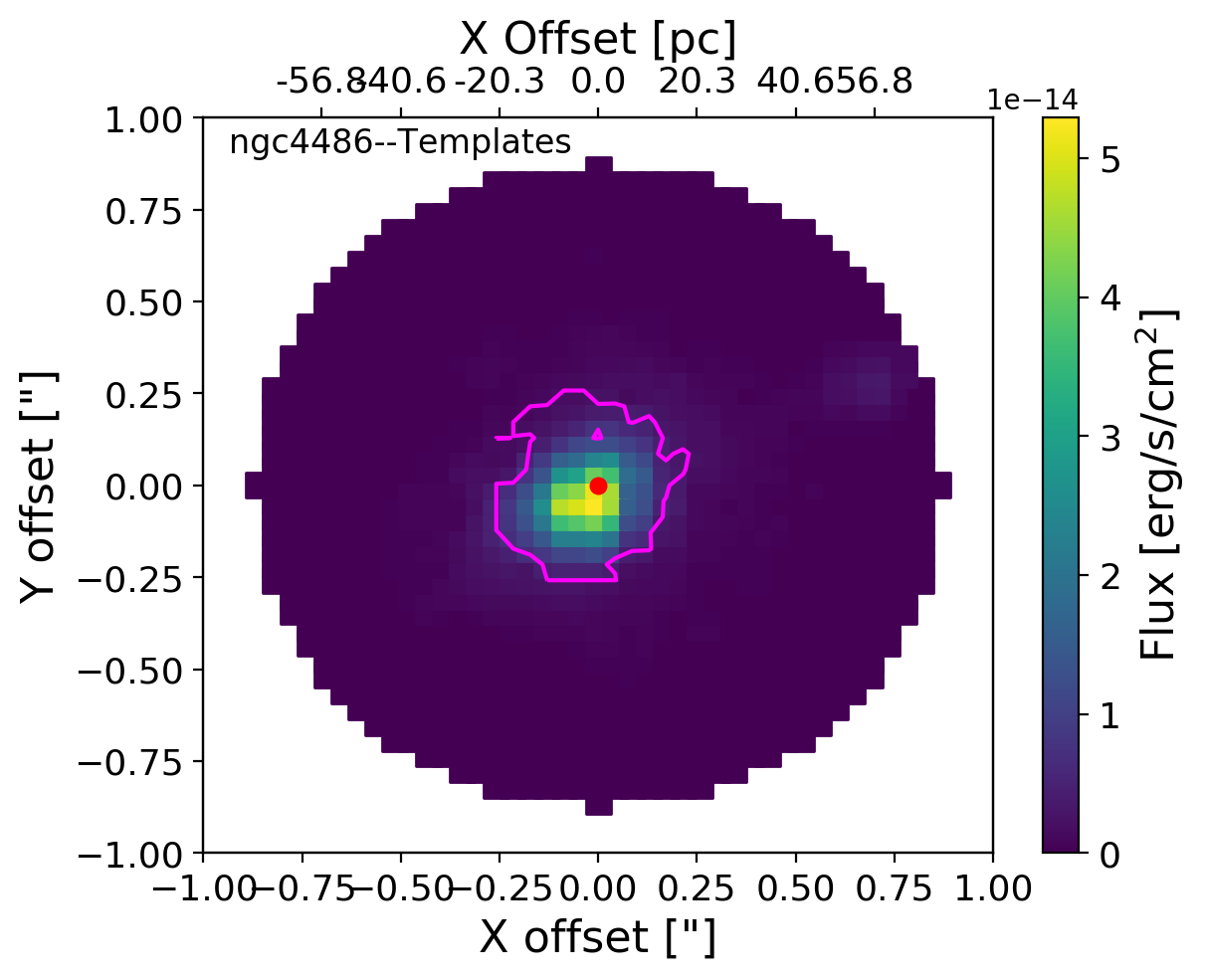}{0.5\textwidth}{(a)}
          \fig{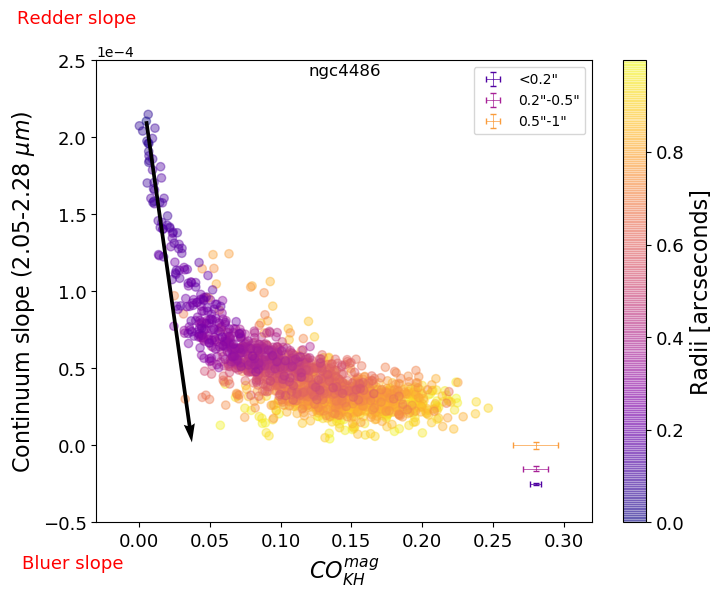}{0.5\textwidth}{(b)}}
\gridline{\fig{ngc4258.png}{0.5\textwidth}{(a)}
          \fig{ngc4258_CO_ew.png}{0.49\textwidth}{(b)}}
\gridline{\fig{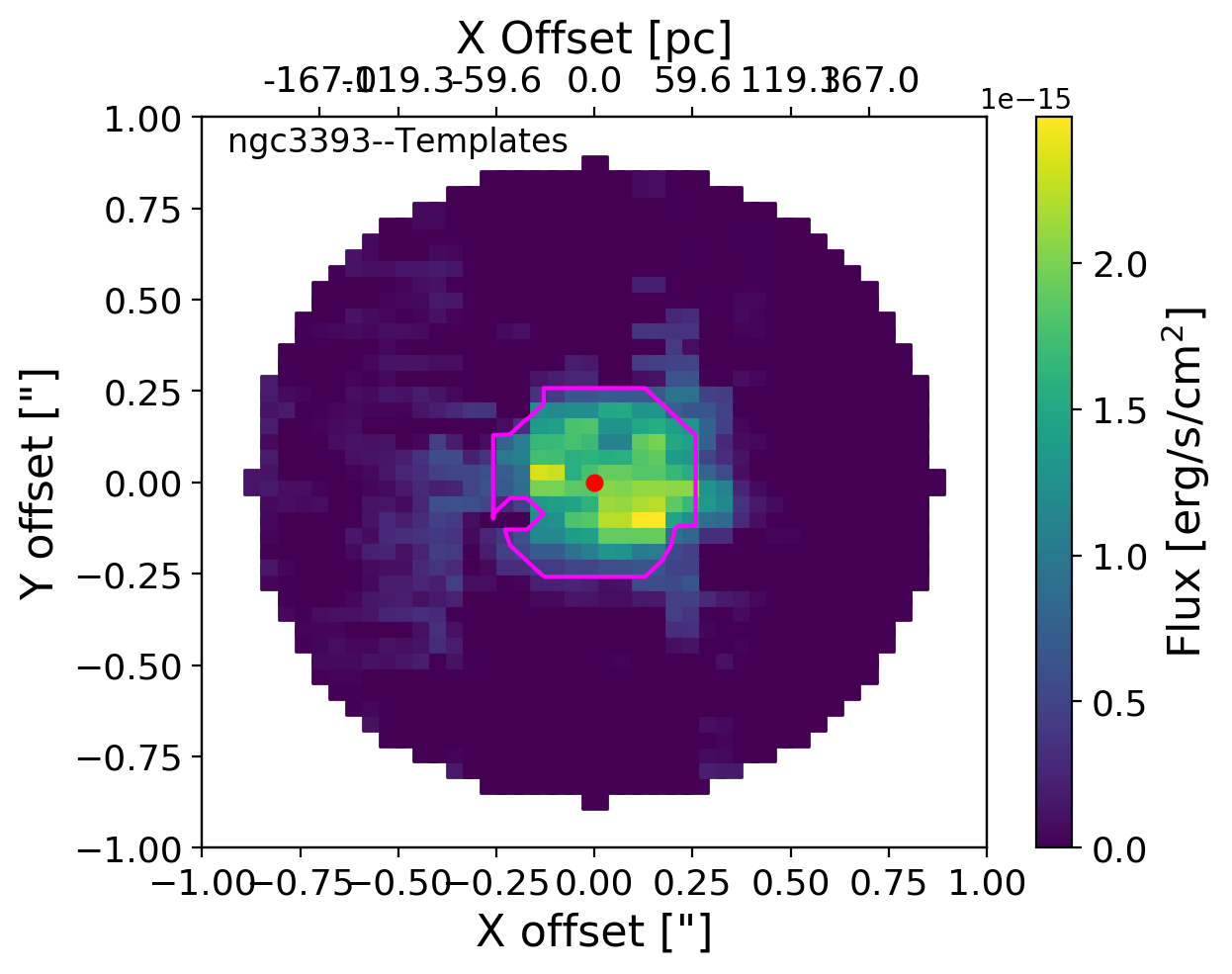}{0.5\textwidth}{(a)}
          \fig{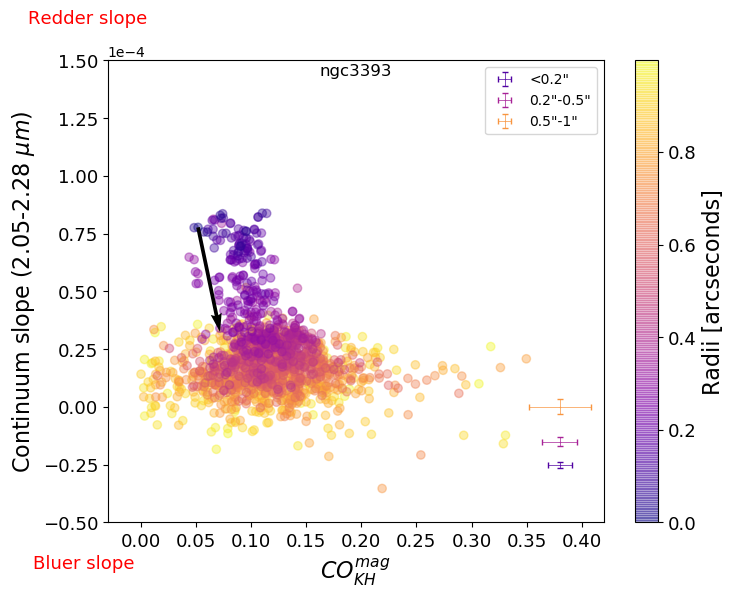}{0.5\textwidth}{(b)}}
\end{figure*}

\begin{figure*}
\gridline{\fig{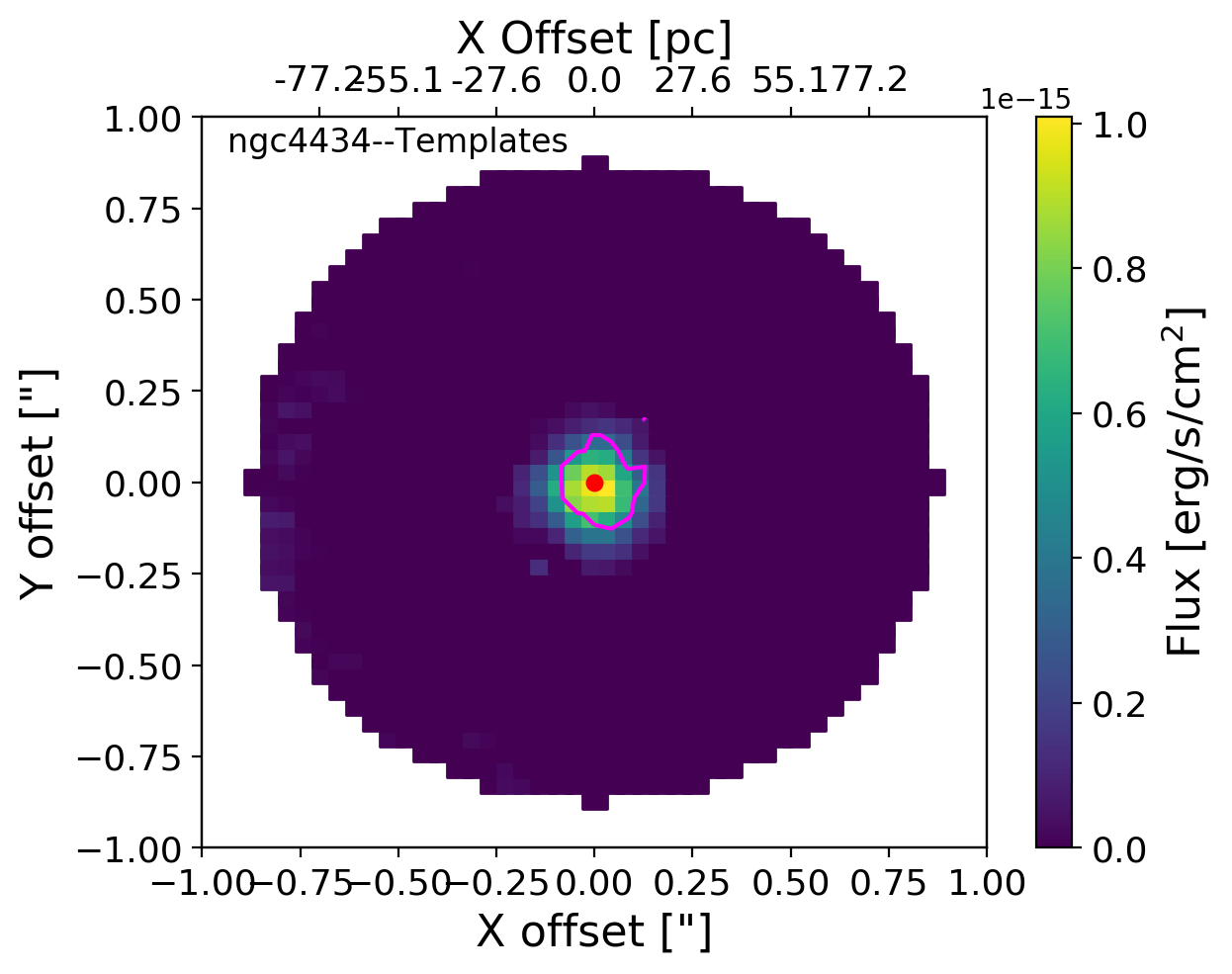}{0.5\textwidth}{(a)}
          \fig{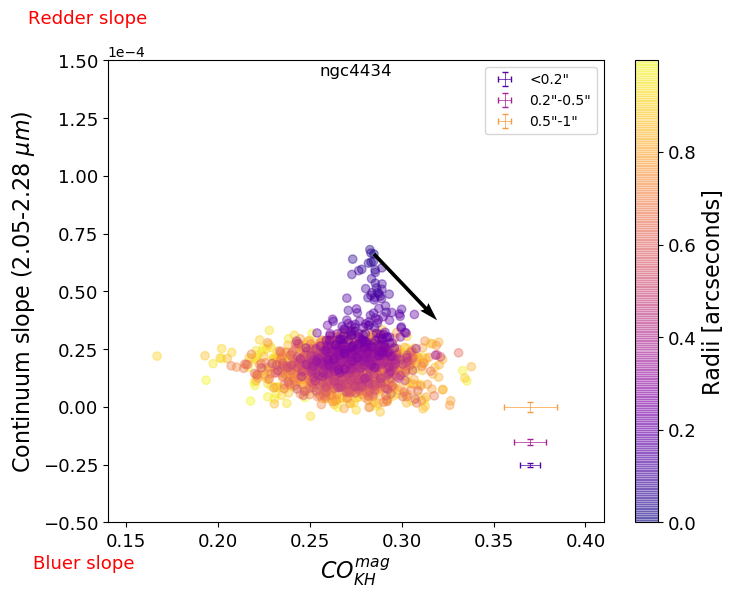}{0.5\textwidth}{(b)}}
\end{figure*}

\begin{figure*}
\plottwo{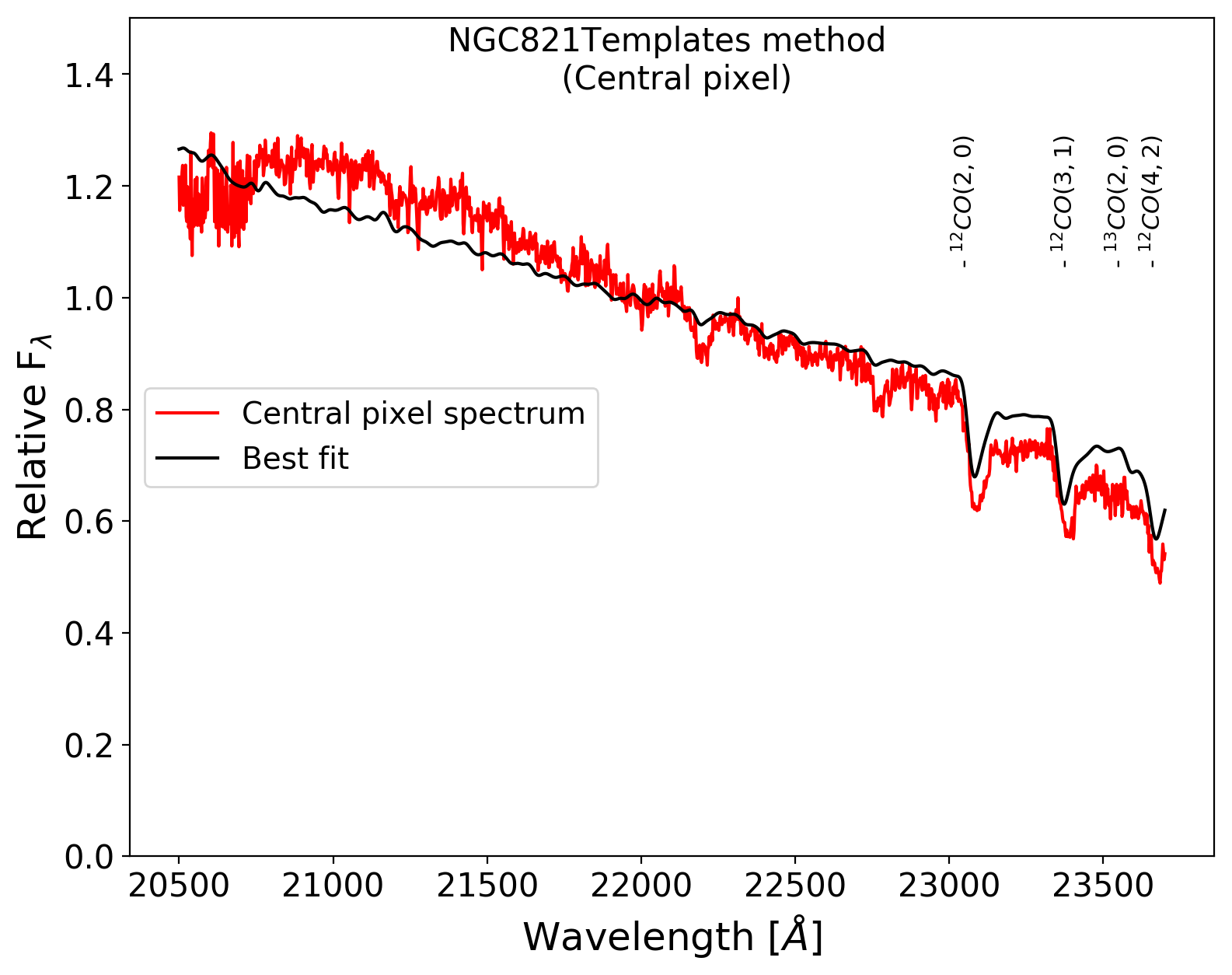}{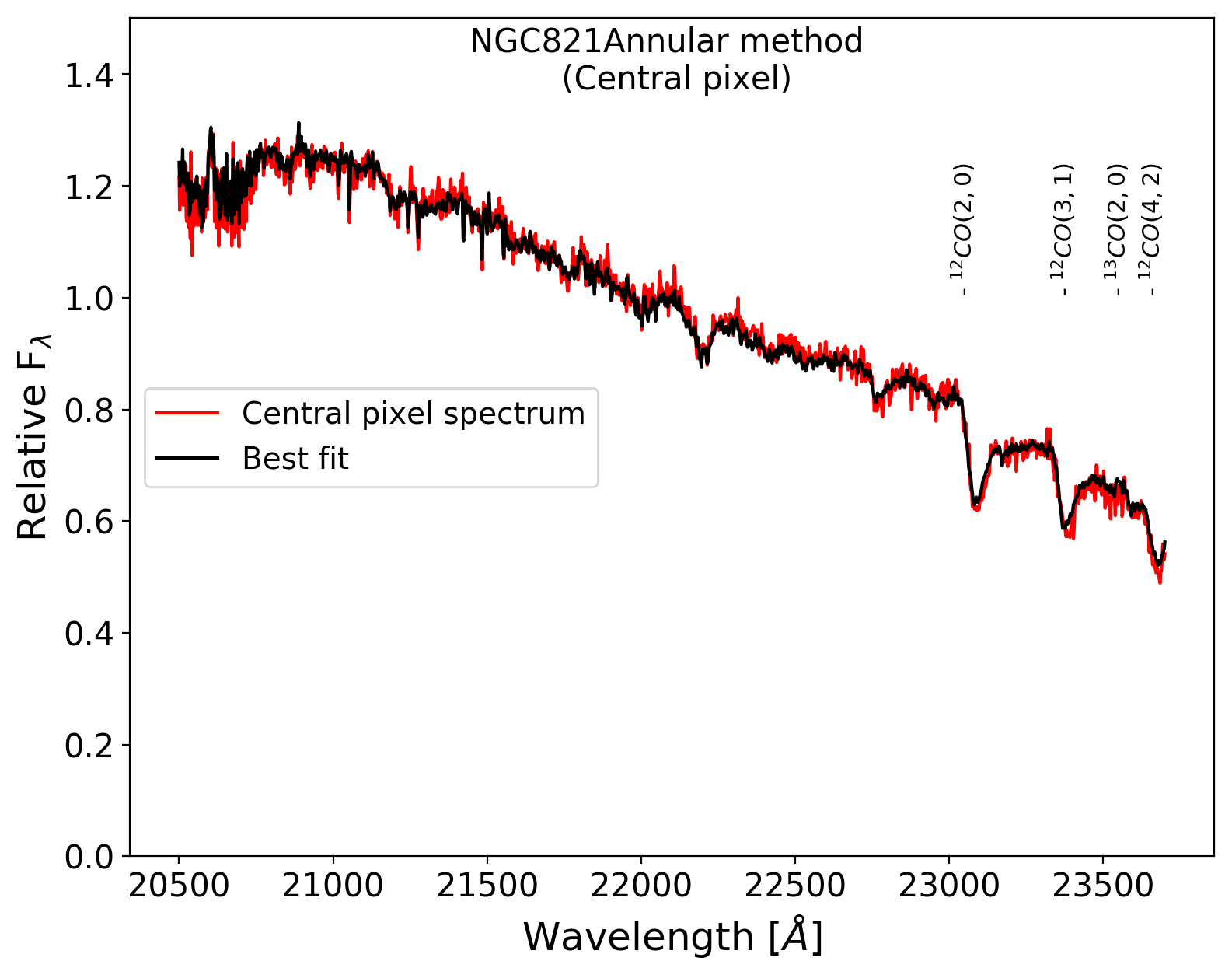}
\caption{Comparison of the best fit for the central pixel of NGC821 with our two different fitting methods described in \ref{subsec:templatesmethod} and \ref{subsec:annularmethod}.  The y-axis represent the relative flux per unit of wavelength. \label{fig:ngc821_comparison}}
\end{figure*}


\end{document}